\documentclass[aps,pra,preprint,groupedaddress,showpacs,preprintnumbers,amsmath,amssymb,superscriptaddress]{revtex4-1}

\usepackage{graphicx}
\usepackage{dcolumn}
\usepackage{bm}

\usepackage{enumerate}
\usepackage{subfigure}
\usepackage{subeqnarray}
\usepackage{cases}
\usepackage[arrow]{hhtensor}     
\usepackage[all]{xy}

\newcommand{\me}{\mathrm{e}}
\newcommand{\mi}{\mathrm{i}}
\newcommand{\dif}{\mathrm{d}}

\begin{document}


\title{Analytical form of light-ray tracing in invisibility cloaks}

\author{Ruo-Yang Zhang}
\email{zhangruoyang@gmail.com}
\affiliation{
 Theoretical Physics Division, Chern Institute of Mathematics, Nankai University, Tianjin,  300071, P.R. China
}

\author{Qing Zhao}%
\email{qzhaoyuping@bit.edu.cn}
\affiliation{
Department of Physics, College of Science, Beijing Institute of Technology, Beijing, 100081, P.R. China
}

\author{Mo-Lin Ge}%
\email{geml@nankai.edu.cn}
\affiliation{
 Theoretical Physics Division, Chern Institute of Mathematics, Nankai University, Tianjin,  300071, P.R. China
}
\affiliation{
Department of Physics, College of Science, Beijing Institute of Technology, Beijing, 100081, P.R. China
}


\begin{abstract}
In this paper, we review the methodology of transformation optics, which can construct invisibility cloak through the transformation of coordinates based on the form invariance of Maxwell's equations. Three different ways to define the components of electromagnetic fields are compared for removing some ambiguities. The analytical expressions of light-ray and wave-normal ray are derived in spherical and cylindrical ideal invisibility cloaks created with any continuous radial transformation functions, and their physical interpretation is also given. Using the duality principle in anisotropic media, we prove that light-ray vector satisfies ``ray-vector eikonal equation'' corresponding to the usual ``wave-vector eikonal equation''. The results interpret why the wave vector maps to the ray vector transferring from the virtual space to the physical space, but not the wave vector. As an application, we investigate the special transformation functions which make the light-ray function satisfy harmonic equation.
\end{abstract}


\maketitle \setlength{\baselineskip}{18pt}

\section{Introduction}

 Recently, invisibility cloak has attracted much attention and widely study \cite{Pendry2006Sci, Leonhardt2006Sci, Greenleaf2003PM, Schurig2006OE, Leonhardt2006NJP, Milton2006NJP, Leonhardt2009PO, Chen2009JOA, Chen2010NatMat, Cummer2006PRE, Chen2007PRL, Ruan2007PRL, Zhang2007PRB, Luo2008APL, Luo2008PRB, Cai2007APL, Niu2009OE, Yao2008APL, Chen2008JAP}. Its theory is based on the form invariance of 3-dimensional Maxwell's equations under a coordinate transformation \cite{Pendry2006Sci, Schurig2006OE, Leonhardt2006NJP, Milton2006NJP}. Briefly, the property of material, such as $\varepsilon$, $\mu$ in a vacuum Euclidean space with an original orthogonal coordinate, will be transformed into a different form through a coordinate transformation. However, the form of Maxwell's equations is invariant. The transformed $\varepsilon$ and $\mu$ could be alternatively interpreted as a new different set of material properties in the original orthogonal coordinate system. Based on this point, the developments have generated a new subject of transformation optics \cite{Leonhardt2009PO, Chen2009JOA, Chen2010NatMat}. Meanwhile, a conformal mapping method has been applied to design a medium which can achieve invisibility in the geometric limit \cite{Leonhardt2006Sci}. The validation of the invisible cloak designed with transformation optics method has been verified through several ways, including ray tracing approach \cite{Schurig2006OE, Niu2009OE}, full-wave simulations \cite{Cummer2006PRE}, and analysis based on Mie scattering [11-16]. The early studies in the invisibility cloak were mostly based on linear transformation proposed by Pendry {\it et al.} \cite{Pendry2006Sci, Schurig2006OE, Chen2007PRL, Ruan2007PRL, Zhang2007PRB}. Then different transformations have also been discussed \cite{Luo2008APL, Cai2007APL}. Luo  {\it et al.} proved that the permittivity and permeability created with a most general form of transformation can realize spherical invisibility cloaks and give the analytical calculation of electromagnetic fields for any spherical cloak using Mie scattering method \cite{Luo2008PRB}.

In the pioneering work of Pendry {\it et al.}\cite{Pendry2006Sci}, it had been noted that the Poynting vectors acting as the tangent vectors of light should propagate to surround the hidden area in the cloak layer. They used the canonical equations of a double anisotropic material optical system to describe the light propagation \cite{Schurig2006OE}. According to the analogy between the curved space and double anisotropic media, Ref. \cite{Niu2009OE} described the ray traces with the geodesic equation in the virtual curved system, and gave the proof that it is equivalent to canonical equations. Since the original Euclidean space (virtual space) is a vacuum, the Poynting vectors and wave vectors of a beam of light traveling in it are parallel to each other. Whereas, when mapping to physical space where a cloak is placed, Poynting vectors and wave vectors are not paralleling any more, which has been pointed out and discussed by Leonhardt in \cite{Leonhardt2010Dover}. In this paper, however, we will expound this point from a different perspective by using ``ray-vector eikonal equation'' and corresponding relations.

The paper is organized as follows. At the beginning, we compare three common ways to define the components of electromagnetic fields in 3-D curvilinear coordinates, which are named Minkowski's definition, Landau's definition, and the components in normalized bases. To do this is benefit to clarify some ambiguities caused by the mistiness of definition. After that, We repeat the fundamentals of transformation optics, which contain the corresponding relations between the curvilinear coordinate system of virtual space ({\bf S} system) and real physical space ({\bf P} space), and emphasize that the validity of transformation optics is not only based on the form of Maxwell's equations, but also based on the correspondence of boundary conditions between the two systems.

After explication of the fundamentals, the analytical solution of electromagnetic fields is derived in spherical and cylindrical cloaks with arbitrary radial transformation functions. The Poynting vectors and wave vectors are further calculated, and the analytical expressions of light rays and wave-normal rays are obtained, respectively. Then, we apply the expressions to four types of cloak made by four different transformation functions. Furthermore, we provide the physical interpretation to the analytical solution of light-rays, and verify it satisfies the geodesic equation in {\bf S} system discussed in \cite{Niu2009OE}. Through duality principle of anisotropic material, we obtain ``ray-vector eikonal equation'' corresponding to the usual ``wave-vector eikonal equation''. And through the corresponding relations, we prove that the covariant component of wave vector $k_i$ in {\bf S} system maps to wave vector $k_{{\scriptscriptstyle(\mathrm{P})}i}$ in {\bf P} space, however, its contravariant component $k^i$ maps to ray vector $s_{{\scriptscriptstyle(\mathrm{P})}}^i$ in {\bf P} space. This is the precise reason that wave vector and Poynting vector are split in {\bf P} space.

Finally, as an application, we find out the general form of transformation functions that makes the light-ray expression satisfy harmonic equation. It is just the conformal transformation function, which Leonhardt applied to design invisibility device in the geometric limit \cite{Leonhardt2006Sci}. In this case, Poynting vectors and wave vectors have neither divergence nor curl.

\section{Coordinate transformation of electromagnetic quantities}

\subsection{Definition of electromagnetic fields in 3-D orthogonal curvilinear coordinates}

For an arbitrary 3-D curvilinear coordinate system $\{x^i\}$ to describe spatial part of a flat spacetime, its bases $\vec{e}_i=\partial\vec{r}/\partial x^i$ are often called coordinate bases or holonomic bases \cite{Misner1973Freeman}, and its spatial metric is $\gamma_{ij}=\vec{e}_i\cdot\vec{e}_j$. The components of a tensor $\matr{T}$ are defined by $\matr{T}=T^{ij}\vec{e}_i\vec{e}_j$.

The component forms of Maxwell's equations in any 3-D curvilinear coordinate system are:
\begin{subequations}\label{Maxwelleqs3Dcurvilinear}
  \begin{equation}
  \nabla_i \bar{B}^i\ =\ \frac{1}{\sqrt{\gamma}}\partial_i(\sqrt{\gamma}\bar{B}^i)\ =0, \quad\quad\quad  \frac{\partial \bar{B}^i}{\partial t}+\epsilon^{ijk} \partial_j \bar{E}_k\ =\ \frac{\partial \bar{B}^i}{\partial t}+\frac{1}{\sqrt{\gamma}}e^{ijk}\partial_j\bar{E}_k\ =\ 0,
  \end{equation}
  \begin{equation}
  \nabla_i \bar{D}^i\ =\ \frac{1}{\sqrt{\gamma}}\partial_i(\sqrt{\gamma}\bar{D}^i)\ =\rho, \quad  -\frac{\partial \bar{D}^i}{\partial t}+\epsilon^{ijk} \partial_j \bar{H}_k\ =\ -\frac{\partial \bar{D}^i}{\partial t}+\frac{1}{\sqrt{\gamma}}e^{ijk} \partial_j \bar{H}_k\ =\ 0,
  \end{equation}
\end{subequations}
where $\nabla_i$ is 3-D covariant derivative, $\gamma$ is the determinate of $\gamma_{ij}$, $e^{ijk}$ is 3-order completely antisymmetric symbol, and $\epsilon^{ijk}=e^{ijk}/\sqrt{\gamma}$ is 3-order Levi-Civita tensor. We use Latin indices $i, j, \cdots$ to denote 1 to 3 for three spatial components. The components of electromagnetic vectors in these equations are resolved in holonomic bases, and this mode of components is named Landau's definition (see Appendix A for general Landau's definition in 4-D curved spacetime). We add a bar on these components to discriminate Minkowski's definition, which will be introduced below.

In this representation, $\vec{E},\ \vec{H},\ \vec{D},\ \vec{B}$ are all treated as 3-D spatial vectors, which imply their components  satisfy the law of vector  transformation  $T^i=(\partial x^i/\partial x'^{k'})T'^{k'}$, meanwhile $\matr{\varepsilon}$ and $\matr{\mu}$ are treated as 3-D spatial tensors, and their components  satisfy the law of transformation
$\bar{\varepsilon}^{ij} = \Lambda^i_{i'}\Lambda^j_{j'}\;\bar{\varepsilon}'^{i'j'},\ \bar{\mu}^{ij} = \Lambda^i_{i'}\Lambda^j_{j'}\;\bar{\mu}'^{i'j'},$
where $\Lambda^i_{i'}=\partial x^i/\partial x'^{i'}$ is the element of Jacobian matrix. In vacuum, $\bar{\varepsilon}^{ij}/\varepsilon_0=\bar{\mu}^{ij}/\mu_0=\gamma^{ij}$.

In transformation optics, a different way to define electromagnetic fields in curved coordinate is established in order to hold the expressions of Maxwell's equations in Cartesian coordinate \cite{Milton2006NJP, Leonhardt2009PO, Ward1996JMO}. If we define electromagnetic quantities as
\begin{subequations}\label{relationin3D}
  \begin{equation}\label{relationin3D1}
  D^i\ =\ \sqrt{\gamma}\bar{D}^i, \qquad\ B^i\ =\ \sqrt{\gamma}\bar{B}^i,\\[1pt]
  \end{equation}
  \begin{equation}\label{relationin3D2}
  \varepsilon^{ij}=\sqrt{\gamma}\bar{\varepsilon}^{ij},\qquad\ \mu^{ij}=\sqrt{\gamma}\bar{\mu}^{ij}.\\[1pt]
  \end{equation}
  \begin{equation}\label{relationin3D3}
  \hat{\rho}\ =\ \sqrt{\gamma}\rho,\qquad\ \ \ \hat{j}^i\ =\ \sqrt{\gamma}j^i,
  \end{equation}
\end{subequations}
substituting them into Eqs. (\ref{Maxwelleqs3Dcurvilinear}), the
form of Maxwell's equations in curved coordinates changes to the
form appearing in Cartesian coordinates. As a result, the form of
Maxwell's equations expressed by ordinary derivative is invariant
from coordinate transformation. We name this mode of definition as Minkowski's
definition (see Appendix A for general Minkowski's definitions).

Note that the above components $D^i,\ B^i,\ \varepsilon^{ij},\ \mu^{ij}$ are
not the real components of the corresponding vectors or tensors in
the curvilinear coordinate, but they can be regarded as the
components of $\sqrt{\gamma}\vec{D},\ \sqrt{\gamma}\vec{B},\
\sqrt{\gamma}\matr{\varepsilon},\ \sqrt{\gamma}\matr{\mu}$
respectively. These quantities are all vector or tensor density,
which satisfy the law of transformation \cite{Schurig2006OE}
\begin{subequations}\label{tansformation of fields}
  \begin{equation}
  D^i\ =\ \left|\det\left( \Lambda^i_{i'}\right)^{-1}\right|\Lambda^i_{i'}\ D'^{i'},\qquad B^i\ =\ \left|\det\left( \Lambda^i_{i'}\right)^{-1}\right|\Lambda^i_{i'}\ B'^{i'},\\[1pt]
  \end{equation}
  \begin{equation}
  \varepsilon^{ij}\ =\ \left|\det\left( \Lambda^i_{i'}\right)^{-1}\right|\Lambda^i_{i'}\Lambda^j_{j'}\;\varepsilon'^{i'j'},\quad \mu^{ij}\ =\ \left|\det\left( \Lambda^i_{i'}\right)^{-1}\right|\Lambda^i_{i'}\Lambda^j_{j'}\;\mu'^{i'j'}.
  \end{equation}
\end{subequations}
In vacuum, the permittivity and permeability have the simple forms as
\begin{equation}\label{constitutive parameters in Minkowski}
\varepsilon^{ij}=\varepsilon_0\sqrt{\gamma}\gamma^{ij},\quad\mu^{ij}=\mu_0\sqrt{\gamma}\gamma^{ij}.
\end{equation}

The components resolved by unit basis are also commonly used to expressing tensors in curved coordinates. The set of normalized bases is defined as $\hat{e}_i=\vec{e}_i/\|\vec{e}_i\|=\vec{e}_i/\sqrt{\gamma_{ii}}$, yet these bases are anholonomic in that we usually can not find a set of coordinates to satisfy $\hat{e}_i=\partial \vec{r}/\partial x^i$. 
The components in orthonormal  bases are also defined by
$\matr{T}=T_{\langle ij \rangle}\,\hat{e}_i\hat{e}_j$, and there is no distinction between covariant and contravariant components when the coordinates are orthogonal. For a 3-D orthogonal curvilinear coordinate system  the set of orthonormal  bases is corresponding to the tetrad in 4-D spacetime (see Appendix A). Since all the components of a vector in unit bases have the same dimension, they act as the measurement of the vector in each direction. In terms of the relation of holonomic and anholonomic bases, we can get
\begin{equation}\label{components in unit bases}
  \varepsilon^{\langle ij \rangle}\ =\ h_ih_j\,\bar{\varepsilon}^{ij}\ =\  \frac{h_ih_j}{\sqrt{\gamma}}\,\varepsilon^{ij},\qquad \mu^{\langle ij \rangle}\ =\ h_ih_j\,\bar{\mu}^{ij}\ =\  \frac{h_ih_j}{\sqrt{\gamma}}\,\mu^{ij},
\end{equation}
where $h_i=\sqrt{\gamma_{ii}}$. For orthogonal coordinates, $ \sqrt{\gamma}=h_1\,h_2\,h_3$ . In Eqs. (\ref{components in unit bases}), we have canceled the Einstein summation convention. For vacuum, $\varepsilon_{\langle ij \rangle}/\varepsilon_0=\mu_{\langle ij \rangle}/\mu_0=\delta_{ij}$.

\subsection{Three coordinate systems in transformation optics}
The method of transformation optics starts from a virtual 3-D flat Euclidean space with medium of vacuum. 
An orthogonal coordinate system (Cartesian or curvilinear coordinate), {\bf S$'$} system, is  employed to describe the virtual space.
In {\bf S$'$}, the coordinates are written as $x^{\,\prime\,i\,'}$
and the metric is $\gamma\,'_{\,i\,'j\,'}$. However, another
coordinate system {\bf S} with coordinates $x^i$ and  metric
$\gamma_{\ ij}$ can be established to describe it. The transformation of
coordinates is $x^i=x^i(x\,'^{\,i\,'})$.

If we investigate electromagnetic fields propagating in the virtual space, Eqs. (\ref{constitutive parameters in Minkowski}) give the Minkowski's form of permittivity and permeability describing vacuum either in {\bf S} or in  {\bf S$'$}. And Eqs. (\ref{tansformation of fields}) present the transformation law of Minkowski's form between {\bf S} and {\bf S$'$} system.

 Due to the invariance of Maxwell's equations under Minkowski's definition, we can alternatively interpret the  Minkowski's form of permittivity and permeability in {\bf S} system as the properties of a real material in physical space ({\bf P} space). The {\bf P} space is also an Euclidean space and described by the same orthogonal coordinates as {\bf S$'$}. We use ${x_{\scriptscriptstyle(\mathrm{P})}}^i$ to denote the coordinates of {\bf P}, and also add the subscript ``$(\mathrm{P})$'' to other physical quantities in {\bf P} space for definitude. Since we employ the same coordinate system as {\bf S$'$} system, the spatial metric ${\gamma_{\scriptscriptstyle(\mathrm{P})}}_{ij}$ of {\bf P} space hold an identical form with $\gamma\,'_{\,i\,'j\,'}$. On the other hand, the Minkowski's form of electromagnetic quantities in {\bf P} space takes the same formula as that in {\bf S} system of virtual space (see Fig. 1).

In previous articles, the {\bf S} system of virtual space is often
called a curved space because of its nontrivial metric. In
mathematics, it is certainly a metric space or manifold. However,
considering its zero curvature, we only treat it as a curvilinear
coordinate system of virtual flat space. In this paper. we only call
a system space when they have a real physical background of
spacetime, such as virtual space and {\bf P} space. But for the
coordinate systems used to describe a space, we only call them
system but not space. Nevertheless, we do not distinguish the {\bf
P} space itself and its coordinate system, because we only set one
system to describe it.

The corresponding relations of electromagnetic quantities between
physical space and {\bf S} system are shown in Table \ref{Table:
Corresponding relation}. There are several points to
emphasize. First, corresponding means the Minkowski's components of
these quantities regarded as functions of coordinates have same
mathematical forms in the two systems. Second, for an individual
quantity, only the covariant (or contravariant) components hold the
same form in the two systems, yet the other kind of components isn't
congruent. For instance,
$D_{\scriptscriptstyle(\mathrm{P})}^i(x_{\scriptscriptstyle(\mathrm{P})}^i)$
and $D^i(x^i)$ have the same mathematical form, but the forms of
${D_{\scriptscriptstyle(\mathrm{P})}}_i(x_{\scriptscriptstyle(\mathrm{P})}^i)$
and $D_i(x^i)$ are different. Third, the identity is only valid for
the components defined in Minkowski's method, however the components
in Landau's definition and in the unit bases often take the different
form in the two systems, unless the three kinds of components reduce
to equivalent under Cartesian coordinate of physical space.

\begin{figure}[t]
\begin{center}
\includegraphics[width=0.5\columnwidth,clip]{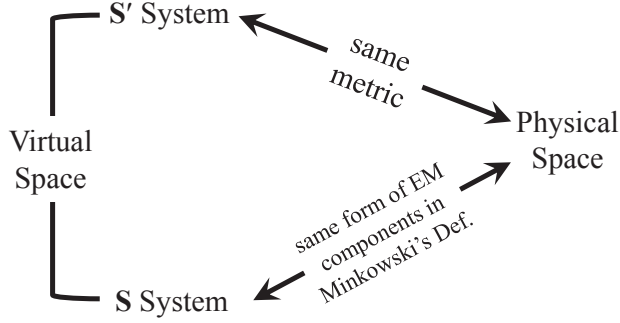}  
\par\end{center}
  \caption{Relations among Physical space, {\bf S$'$} coordinate system and {\bf S} coordinate system of virtual space.  }
\end{figure}

\setlength{\tabcolsep}{12pt}
\renewcommand{\arraystretch}{1.3}
\begin{table}[h]
\setlength{\belowcaptionskip}{5pt}
\centering\caption{Corresponding relations.}
\begin{tabular}{p{65pt}cccccc}
\hline
      &      \multicolumn{6}{c}{Electromagnetic quantities }    \\
\cline{2-7}
\multicolumn{1}{r}{Physical space : } \rule[-8pt]{0pt}{25pt}&     $D_{\scriptscriptstyle(\mathrm{P})}^i$  & ${E_{\scriptscriptstyle(\mathrm{P})}}_i$    &  $B_{\scriptscriptstyle(\mathrm{P})}^i$  &  ${H_{\scriptscriptstyle(\mathrm{P})}}_i$  &   ${\varepsilon_{\scriptscriptstyle(\mathrm{P})}}^{ij}$  &   ${\mu_{\scriptscriptstyle(\mathrm{P})}}^{ij}$  \\
\multicolumn{1}{r}{{\bf S}\ \ system : }  \rule[-7pt]{0pt}{15pt}   &  $D^i$  & $E_i$    &  $B^i$  &  $H_i$  &   $\varepsilon^{ij}$  &   $\mu^{ij}$  \\ \hline
\end{tabular}\label{Table: Corresponding relation}
\end{table}

Now our discussion is restricted on the condition that {\bf S}
system is an orthogonal coordinate system with
$\gamma_{ij}={h_i}^2\delta_{ij}$. Based on the corresponding
relations, Eqs. (\ref{tansformation of fields}), and
(\ref{components in unit bases}), we derive the permittivity and
permeability in physical space:
\begin{equation}\label{permittivity and permeability in P space}
  \varepsilon_{{\scriptscriptstyle(\mathrm{P})}\langle ij \rangle}/\varepsilon_0\ =\  \mu_{{\scriptscriptstyle(\mathrm{P})}\langle ij \rangle}/\mu_0\ =\ \sum_{i'}\frac{h'_1\,h'_2\,h'_3}{h_{{\scriptscriptstyle(\mathrm{P})}1}\,h_{{\scriptscriptstyle(\mathrm{P})}2}\,h_{{\scriptscriptstyle(\mathrm{P})}3}}
  \frac{h_{{\scriptscriptstyle(\mathrm{P})}i}\,h_{{\scriptscriptstyle(\mathrm{P})}j}}{(h'_{i'})^2}\left|\det\left( \Lambda^i_{i'}\right)^{-1}\right|\Lambda^i_{i'}\Lambda^j_{i'}.
\end{equation}
Here we take the transformation matrix as a function of $x^i$ and
$x'^{i'}$, $ i.e.$ $\Lambda^i_{i'}=\Lambda^i_{i'}(x^i,\;x'^{i'})$,
meanwhile the independent variables should be
$x_{\scriptscriptstyle(\mathrm{P})}^i$ and $x'^{i'}$ in the above
formula.

For spherical invisibility cloak, if we select {\bf S$'$} system as
spherical coordinate system, define the inverse transformation from
{\bf S$'$} to {\bf S} system as $r'=f(r),\ \theta'=\theta,\
\phi'=\phi$, and insert it into Eq. (\ref{permittivity and
permeability in P space}), the constitutive parameters become
\cite{Luo2008PRB}
\begin{equation}\label{constitutiveparametersinsphericalcloak}
  \big( \varepsilon_{{\scriptscriptstyle(\mathrm{P})}\langle ij \rangle}/\varepsilon_0 \big)\ = \ \big( \mu_{{\scriptscriptstyle(\mathrm{P})}\langle ij \rangle}/\mu_0 \big)\ =\
  \mathrm{diag}\left(\frac{f^2(r_{\scriptscriptstyle(\mathrm{P})})}{{r_{\scriptscriptstyle(\mathrm{P})}}^2f'(r_{\scriptscriptstyle(\mathrm{P})})},\ \  f'(r_{\scriptscriptstyle(\mathrm{P})}), \ \ f'(r_{\scriptscriptstyle(\mathrm{P})})
  \right).
\end{equation}
For arbitrary transformation function, it can be completely
invisible, as long as $f(a)=0$ and $f(b)=b$,where  $a,\ b$ are the
inner and outer radii of the cloak layer, even if $f(r)$ is
continuous but non-differentiable \cite{Luo2008PRB}. Similarly, the
constitutive parameters for cylindrical cloak obtained by the mere
radial transformation $r'=f(r),\ \theta'=\theta,\ z'=z$ are
\cite{Cai2007APL}
\begin{equation}\label{constitutiveparametersincylindricalcloak}
  \big( \varepsilon_{{\scriptscriptstyle(\mathrm{P})}\langle ij \rangle}/\varepsilon_0 \big)\ = \ \big( \mu_{{\scriptscriptstyle(\mathrm{P})}\langle ij \rangle}/\mu_0 \big)\ =\
  \mathrm{diag}\left(\frac{f(r_{\scriptscriptstyle(\mathrm{P})})}{r_{\scriptscriptstyle(\mathrm{P})}f'(r_{\scriptscriptstyle(\mathrm{P})})},\ \ \frac{r_{\scriptscriptstyle(\mathrm{P})}f'(r_{\scriptscriptstyle(\mathrm{P})})}{f(r_{\scriptscriptstyle(\mathrm{P})})},\ \ \frac{f(r_{\scriptscriptstyle(\mathrm{P})})f'(r_{\scriptscriptstyle(\mathrm{P})})}{r_{\scriptscriptstyle(\mathrm{P})}})
  \right),
\end{equation}
where the components are described by cylindrical coordinates. The
arbitrary $f(r)$ will achieve completely invisible only if it
satisfies the boundary condition $f(a)=0$ and $f(b)=b$ ( see the
proof in Appendix B).

\subsection{Boundary conditions for perfect invisibility cloak }
The congruity of Maxwell's equations with the corresponding
relations between {\bf S} system and {\bf P} space is regarded as
the foundation of transformation optics. However, despite of the
existence on the same form of Maxwell's equations, they won't give
the same solution unless the boundary conditions of the two systems
are also identical.

For example, the realization of spherical and cylindrical
invisibility cloak requires the inverse transformation function
satisfies $f(a)=0$ and $f(b)=b$. The $f(a)=0$ causes the domain
$r'<a$ in {\bf S$'$} system to shrink into its original point in
{\bf S}, so that the coordinates with radial component $r<a$ would
never appear in the expression of electromagnetic fields in {\bf S}
system. It leads to decoupling of fields between the hidden region
$r_{\scriptscriptstyle(\mathrm{P})}<a$ and cloak layer in the {\bf
P} space. Pendry {\it et al.} pointed out that the condition
$f(b)=b$ makes the constitutive parameters satisfy perfectly matched
layer (PML) conditions
$\varepsilon_{{\scriptscriptstyle(\mathrm{P})}\langle \theta\theta
\rangle}=\varepsilon_{{\scriptscriptstyle(\mathrm{P})}\langle
\phi\phi
\rangle}=1/\varepsilon_{{\scriptscriptstyle(\mathrm{P})}\langle rr
\rangle}$ and $\mu_{{\scriptscriptstyle(\mathrm{P})}\langle
\theta\theta \rangle}=\mu_{{\scriptscriptstyle(\mathrm{P})}\langle
\phi\phi \rangle}=1/\mu_{{\scriptscriptstyle(\mathrm{P})}\langle rr
\rangle}$ at the outer surface of the cloak
($r_{\scriptscriptstyle(\mathrm{P})}=b$)\cite{Pendry2006Sci},
therefore the cloak is reflectiveness. However there is a more
profound reason produced by the corresponding relation of the boundary
conditions between { \bf S$'$} system and {\bf P} space.

First, we consider a spherical cloak designed by general transformation $x^{i}(r',\theta',\phi')$ from {\bf S$'$} to {\bf S} system. In {\bf P} space, the boundary conditions at the interface $r=b$ are
\begin{equation}\label{boundary conditions}
  (\vec{E}_{\scriptscriptstyle(\mathrm{P})}\big|_{r=b^+}-\vec{E}_{\scriptscriptstyle(\mathrm{P})}\big|_{r=b^-})\times\hat{e}_r\ =\ 0,\qquad (\vec{D}_{\scriptscriptstyle(\mathrm{P})}\big|_{r=b^+}-\vec{D}_{\scriptscriptstyle(\mathrm{P})}\big|_{r=b^-})\cdot\hat{e}_r\ =\ 0,
\end{equation}
where $\hat{e}_r$ acts as the normal vector of interface. The
boundary conditions indicate the tangential component of $\vec{E}$
and normal component of $\vec{D}$ are continuous across the
interface, and the two boundary conditions can completely determine
$\vec{E}_{\scriptscriptstyle(\mathrm{P})}|_{r=b^-}$,  if
$\vec{E}_{\scriptscriptstyle(\mathrm{P})}|_{r=b^+}$ is given. Since
the metric of {\bf P} space is continuous, the equality of
components is always tenable no matter what definitions. Thus, we
have the continuous relations in Minkowski's form
\begin{equation}\label{boundary conditions in Minkowski definition}
  E_{{\scriptscriptstyle(\mathrm{P})}\theta}\big|_{r=b^+} = E_{{\scriptscriptstyle(\mathrm{P})}\theta}\big|_{r=b^-},\quad E_{{\scriptscriptstyle(\mathrm{P})}\phi}\big|_{r=b^+} = E_{{\scriptscriptstyle(\mathrm{P})}\phi}\big|_{r=b^-},\quad  D_{\scriptscriptstyle(\mathrm{P})}^r\big|_{r=b^+} = D_{\scriptscriptstyle(\mathrm{P})}^r\big|_{r=b^-}.
\end{equation}
Non-reflection means the electric field outside the cloak only has
the incident wave. According to the corresponding relations in Table
\ref{Table: Corresponding relation}, the electric field in {\bf S}
system should also satisfy the relations
\begin{subequations}\label{boundary conditions in S}
\begin{equation}
  E'_{\theta'}\big|_{r'=b^+} = E_\theta\big|_{r=b^+} = E_\theta\big|_{r=b^-} = \Lambda^{i'}_\theta\, E'_{i'}\big|_{x'^{i'}(b^-,\theta,\phi)},
\end{equation}
\begin{equation}
  E'_{\phi'}\big|_{r'=b^+} = E_\phi\big|_{r=b^+} = E_\phi\big|_{r=b^-} = \Lambda^{i'}_\phi\, E'_{i'}\big|_{x'^{i'}(b^-,\theta,\phi)},
\end{equation}
\begin{equation}
  D'^{r'}\big|_{r'=b^+} = D^r\big|_{r=b^+} = D^r\big|_{r=b^-} = \left|\det\left( \Lambda^i_{i'}\right)^{-1}\right|\Lambda^{r}_{i'}\, D'^{i'}\big|_{x'^{i'}(b^-,\theta,\phi)},
\end{equation}
\end{subequations}
where 
$E'_{\theta'}|_{r'=b^+} = E_\theta|_{r=b^+}$ since {\bf S$'$} and {\bf S} system hold the same coordinates in the domain $r>b$. Note that the boundary conditions Eqs. (\ref{boundary conditions}) certainly satisfy in virtual space. 
However the metric is discontinuous at the interface $r=b$ in {\bf
S} system, therefore the relation of components $E_\theta|_{r=b^+} =
E_\theta|_{r=b^-}$ is not trivial, but a strong limit to the metric.

By comparing the first and last term, we note that the electric field at $r'=b$ is demanded to have a
fixed relation to the field at $x'^{i'}(b^-,\theta,\phi)$ in {\bf
S$'$} system. Because of the locality of electromagnetic fields,  a
certain relation of fields between different positions can not be
established. Therefore $(b,\, \theta,\, \phi)$ and
$x'^{i'}(b^-,\theta,\phi)$ must be the same point:
\begin{equation}
  r'(b^-,\theta,\phi)\, =\, b,\quad \theta'(b^-,\theta,\phi)\,=\, \theta,\quad \phi'(b^-,\theta,\phi)\,=\, \phi.
\end{equation}
This is the requirement of non-reflection. Thus, we have
\begin{displaymath}
  \left(\frac{\partial \theta'}{\partial \theta}\,=\, \frac{\partial \phi'}{\partial \phi}\right)\Big|_{r=b^-}=\, 1,\ \left(\frac{\partial r'}{\partial \theta}\, =\, \frac{\partial r'}{\partial \phi}\, =\, \frac{\partial \theta'}{\partial \phi} =\, \frac{\partial \phi'}{\partial \theta}\right)\Big|_{r=b^-} =\,0.
\end{displaymath}
Substituting it into Eqs. (\ref{boundary conditions in S}), the
equalities are directly satisfied. A similar study to the boundary
conditions of magnetic field can give the identical result.

In fact, the interface at $r=b$ in {\bf S} system generally doesn't correspond
to the interface $r'=b$ in {\bf S$'$} system for an arbitrary
transformation. Thus the metric is usually discontinuous across the
interface, and we even hardly measure the element of length on the
interface. For example, if the inverse transformation is simply
$r'=f(r),\ \theta'=\theta,\ \phi'=\phi$ as $r<b$, the elements of
length on the inner and outer surface of the interface are
respectively
\begin{displaymath}
  \dif l^2\big|_{r=b^-} =\ f(b)^2\,\big(\dif \theta^2\, +\, \sin^2\theta\dif \phi^2\big),\quad  \dif l^2\big|_{r=b^+}=\ b^2\,\big(\dif \theta^2\, +\, b^2\sin^2\theta\dif \phi^2\big).
\end{displaymath}
Nothing but $f(b)=b$ gives $\dif l^2|_{r=b^-}=\dif l^2|_{r=b^+}$ with arbitrary $\dif \theta$ and $\dif \phi$. It is also available for cylindrical cloak by a similar analysis that non-reflection requires invariance of coordinates on the interface  before and after transformation.

\section{Analytical expressions of light-ray and wave-normal ray}

\subsection{Spherical cloak}
Considering a spherical cloak, the cloak hides a homogenous-media
sphere, and its center is at the origin of a spherical coordinate
system. A plane wave
$\vec{E}^{\mathrm{in}}_{\scriptscriptstyle(\mathrm{P})}=E_0\me^{\mi
k_0r\cos\theta}\hat{e}_x$ with $x$ polarization is incident to the
cloak. It has been proved that both the fields
$\vec{E}^{\mathrm{int}}_{\scriptscriptstyle(\mathrm{P})}$ inside the
internal area ($r<a$), and the scattering fields
$\vec{E}^{\mathrm{sc}}_{\scriptscriptstyle(\mathrm{P})}$ outside
cloak ($r>b$) varnish if the constitutive parameters of cloak
satisfy the formula Eq.
(\ref{constitutiveparametersinsphericalcloak}) and the boundary
conditions $f(a)=0,\ f(b)=b$. Meanwhile the electromagnetic fields
in cloak layer ($a<r<b$) have been given in \cite{Luo2008PRB}
\begin{subequations}\label{Fieldsinsphericalcloak}
  \begin{flalign}
  \vec{E}^\mathrm{c}_{\scriptscriptstyle(\mathrm{P})} & =  E_0\left( f'(r)\sin\theta\cos\phi\hat{e}_r+\frac{f(r)}{r}\cos\theta\cos\phi\hat{e}_\theta-\frac{f(r)}{r}\sin\phi\hat{e}_\phi \right)\me^{\mi k_0f(r)\cos\theta},\\
  \vec{D}^\mathrm{c}_{\scriptscriptstyle(\mathrm{P})} & =  \varepsilon_0E_0\frac{f(r)}{r}\left( \frac{f(r)}{r}\sin\theta\cos\phi\hat{e}_r+f'(r)\cos\theta\cos\phi\hat{e}_\theta-f'(r)\sin\phi\hat{e}_\phi \right)\me^{\mi k_0f(r)\cos\theta},\\
  \vec{H}^\mathrm{c}_{\scriptscriptstyle(\mathrm{P})} & =  \frac{E_0}{\eta_0}\left( f'(r)\sin\theta\sin\phi\hat{e}_r+\frac{f(r)}{r}\cos\theta\sin\phi\hat{e}_\theta+\frac{f(r)}{r}\cos\phi\hat{e}_\phi \right)\me^{\mi k_0f(r)\cos\theta},\\
  \vec{B}^\mathrm{c}_{\scriptscriptstyle(\mathrm{P})} & =  \frac{E_0f(r)}{c r}\left( \frac{f(r)}{r}\sin\theta\sin\phi\hat{e}_r+f'(r)\cos\theta\sin\phi\hat{e}_\theta+f'(r)\cos\phi\hat{e}_\phi \right)\me^{\mi k_0f(r)\cos\theta},
\end{flalign}
\end{subequations}
where $\eta_0=\sqrt{\mu_0/\varepsilon_0}$ and $k_0=\omega/c$ are
the impedance and wave number of vacuum respectively. Since only fields in {\bf
P} space are discussed in this section, we omit the subscript
``($\mathrm{P}$)'' in all the expressions of coordinate, but still
retain it in the expressions of electromagnetic quantities.
According to above fields , we obtain the time-averaged Poynting
vector
\begin{equation}\label{Poyntinginsphericalcloak}
  \langle \vec{S}^\mathrm{c}_{\scriptscriptstyle(\mathrm{P})}\rangle\ =\ \frac{1}{2}\mathrm{Re}(\vec{E}^\mathrm{c}_{\scriptscriptstyle(\mathrm{P})}\times\vec{H}^\mathrm{c}_{\scriptscriptstyle(\mathrm{P})})\ =\ \frac{{E_0}^2f(r)}{2\eta_0r}\left( \frac{f(r)}{r}\cos\theta\hat{e}_r-f'(r)\sin\theta\hat{e}_\theta \right),
\end{equation}
It can be verified that the Poynting vector satisfies energy
conservation $ \nabla\cdot\langle
\vec{S}^\mathrm{c}_{\scriptscriptstyle(\mathrm{P})}\rangle=0 $
\cite{Born1999Cambridge}. Supposing the medium of cloak is
non-dispersive, the time-averaged energy density of electromagnetic
fields has the simple form
\begin{equation}\label{energydensityinsphericalcloak}
  \langle W^\mathrm{c}_{\scriptscriptstyle(\mathrm{P})}\rangle\ =\ \frac{1}{2}\mathrm{Re}\left(\frac{1}{2}\vec{E}^\mathrm{c}_{\scriptscriptstyle(\mathrm{P})}\cdot\vec{D}^\mathrm{c}_{\scriptscriptstyle(\mathrm{P})}
  +\frac{1}{2}\vec{H}^\mathrm{c}_{\scriptscriptstyle(\mathrm{P})}\cdot\vec{B}^\mathrm{c}_{\scriptscriptstyle(\mathrm{P})}\right)\ =\ \frac{\varepsilon_0{E_0}^2f(r)^2f'(r)}{2r^2}.
\end{equation}
Thus, the ray (energy) velocity in cloak is given as \cite{Born1999Cambridge, Chen1985McGraw}
\begin{equation}\label{rayvelocityinsphericalcloak}
  \vec{v}_{{\scriptscriptstyle(\mathrm{P})}ray} =\ \frac{\langle \vec{S}^\mathrm{c}_{\scriptscriptstyle(\mathrm{P})}\rangle}{\langle W^\mathrm{c}_{\scriptscriptstyle(\mathrm{P})}\rangle}\ =\ c\left( \frac{1}{f'(r)}\cos\theta\hat{e}_r-\frac{r}{f(r)}\sin\theta\hat{e}_\theta \right).
\end{equation}
The ray vector is \cite{Chen1985McGraw}
\begin{equation}\label{rayvectorinsphericalcloak}
  \vec{s}_{\scriptscriptstyle(\mathrm{P})} =\ \frac{\vec{v}_{{\scriptscriptstyle(\mathrm{P})}ray}}{\omega}\ =\ {k_0}^{-1}\left( \frac{1}{f'(r)}\cos\theta\hat{e}_r-\frac{r}{f(r)}\sin\theta\hat{e}_\theta \right).
\end{equation}

In terms of the eikonal $\psi=k_0f(r)\cos\theta$, the geometric wave-front, $\psi=\mbox{constant}$,  is
\begin{equation}\label{wave-front}
  f(r)\cos\theta\ \ =\ \ \mbox{const}.
\end{equation}
Then, we can get the wave vector
$\vec{k}_{\scriptscriptstyle(\mathrm{P})}$, which is
orthogonal to the wave-front, in the cloak
\begin{equation}\label{wavevectorinsphericalcloak}
  \vec{k}_{\scriptscriptstyle(\mathrm{P})}\ =\ \nabla\psi\ =\ k_0 \left( f'(r)\cos\theta\hat{e}_r-\frac{f(r)}{r}\sin\theta\hat{e}_\theta \right).
\end{equation}

In the anisotropic media, the ray vector ( time-averaged Poynting
vector ) is tangent to light-ray, however the wave vector is tangent to
wave-normal ray, where $ \vec{s} $ and  $\vec{k}$ are generally not
parallel to each other. The parametric equations of light-ray can be
written as
\begin{subequations}
  \begin{flalign}
  \frac{\dif r}{\dif \lambda} & \ \ = \ \ {s_{\scriptscriptstyle(\mathrm{P})}}^r \ \ =\ \ \frac{1}{k_0f'(r)}\cos\theta ,\label{dr/dlambda_Poynting}\\
  \frac{\dif \theta}{\dif \lambda} & \ \ =\ \ {s_{\scriptscriptstyle(\mathrm{P})}}^\theta \ \ =\ \ -\frac{1}{k_0f(r)}\sin\theta,\label{dtheta/dlambda_Poynting}\\
  \frac{\dif \phi}{\dif \lambda} & \ \ = \ \ {s_{\scriptscriptstyle(\mathrm{P})}}^\phi \ \ =\ \ 0,\label{dphi/dlambda_Poynting}
  \end{flalign}
\end{subequations}
where $\lambda$ is the parameter of light-ray parametric equation,
and the components of ray vector employed above are contravariant in
holonomic spherical coordinate bases. The $\phi$ maintains its initial
value $\phi_0$ through out a light-ray in terms of Eq.
(\ref{dphi/dlambda_Poynting}). By eliminating $\lambda$ from Eq.
(\ref{dr/dlambda_Poynting}) and Eq. (\ref{dtheta/dlambda_Poynting})
, we have
\begin{displaymath}
  \int{ \frac{f'(r)}{f(r)}\ \dif r}\ \ =\ \ -\int \frac{\cos\theta}{\sin\theta}\ \dif \theta.
\end{displaymath}
Then the analytical expression of light-ray is obtained
\begin{equation}\label{lightrayinsphericalcloak}
  f(r)\sin\theta\ \ =\ \ \mbox{const},\qquad \phi\ \ =\ \ \phi_0.
\end{equation}
Here, the integral constant and $\phi_0$ depend on the incident
point on the outer surface of the cloak.

A similar analysis is used to get the parametric equations for the
wave-normal ray
\begin{subequations}
  \begin{flalign}
  \frac{\dif r}{\dif \lambda} & \ \ = \ \ {k_{\scriptscriptstyle(\mathrm{P})}}^r \ \ =\ \ k_0f'(r)\cos\theta ,\label{dr/dlambda_Wavevector}\\
  \frac{\dif \theta}{\dif \lambda} & \ \ =\ \ {k_{\scriptscriptstyle(\mathrm{P})}}^\theta \ \ =\ \ -k_0\frac{f(r)}{r^2}\sin\theta,\label{dtheta/dlambda_Wavevector}\\
  \frac{\dif \phi}{\dif \lambda} & \ \ = \ \ {k_{\scriptscriptstyle(\mathrm{P})}}^\phi \ \ =\ \ 0.\label{dphi/dlambda_Wavevector}
  \end{flalign}
\end{subequations}
Thus, the wave-normal ray equation is
\begin{equation}\label{wave-normalinsphericalcloak}
   \exp\left[\int\frac{f(r)}{r^2f'(r)}\dif r\right]\sin\theta\ \ =\ \ \mbox{const},\qquad \phi\ \ =\ \ \phi_0.
\end{equation}

According to Eq. (\ref{rayvelocityinsphericalcloak}), when $r$
closes to $a$, $f(r)$ tends to 0, and the modulus of ray velocity
increases beyond vacuum light speed $c$ and tends to infinity which
is against the law of causality. The contradiction indicates that
the assumption of non-dispersive cloak is unpractical. In the
presence of dispersion, the time-averaged energy density has a more
complicated formula. It includes the derivative of constitutive
parameters with respect to $\omega$ \cite{Chen1985McGraw}, and then the
superluminal result should disappear. More detailed discussion to
this point can be found in \cite{Yao2008APL, Chen2008JAP}.
Nevertheless, the expressions in Eqs.
(\ref{rayvelocityinsphericalcloak}, \ref{rayvectorinsphericalcloak})
still exactly represent the direction of the ``real ray velocity'',
therefore we will retain the expressions of
$\vec{v}_{{\scriptscriptstyle(\mathrm{P})}ray}$ and
$\vec{s}_{\scriptscriptstyle(\mathrm{P})}$, but limit them only to
stand for the tangent vector of light-ray.

\subsection{Cylindrical cloak}
Considering a cylindrical cloak with indefinite length, its axis is
$z$ axis of a cylindrical coordinate system. The incident wave is a
transverse-electric(TE) wave
$\vec{E}^{\mathrm{in}}_{\scriptscriptstyle(\mathrm{P})}=E_0\me^{\mi
k_0r\cos\theta}\hat{e}_z$ with $z$ polarization propagating towards
$x$ direction. If the cloak hold the perfect invisible parameters
expressed in Eq. (\ref{constitutiveparametersincylindricalcloak}),
the calculation in Appendix B gives
$\vec{E}^{\mathrm{int}}_{\scriptscriptstyle(\mathrm{P})}=0$ in the
inner hidden area ($r<a$), and the scattering
$\vec{E}^{\mathrm{sc}}_{\scriptscriptstyle(\mathrm{P})}=0$ outside
the cloak ($r>b$). Meanwhile the fields in the cloak layer are
\begin{subequations}\label{FieldsinCylindricalcloak}
  \begin{flalign}
  \vec{E}^\mathrm{c}_{\scriptscriptstyle(\mathrm{P})} & =  E_0\me^{\mi k_0f(r)\cos\theta}\hat{e}_z,\\
  \vec{D}^\mathrm{c}_{\scriptscriptstyle(\mathrm{P})} & =  \varepsilon_0E_0\frac{f'(r)f(r)}{r}\me^{\mi k_0f(r)\cos\theta}\hat{e}_z,\\
  \vec{H}^\mathrm{c}_{\scriptscriptstyle(\mathrm{P})} & =  -\frac{E_0}{\eta_0}\left( f'(r)\sin\theta\hat{e}_r+\frac{f(r)}{r}\cos\theta\hat{e}_\theta \right)\me^{\mi k_0f(r)\cos\theta},\\
  \vec{B}^\mathrm{c}_{\scriptscriptstyle(\mathrm{P})} & =  -\frac{E_0}{c}\left( \frac{f(r)}{r}\sin\theta\hat{e}_r+f'(r)\cos\theta\hat{e}_\theta \right)\me^{\mi k_0f(r)\cos\theta}.
\end{flalign}
\end{subequations}
Thus, the time-averaged Poynting vector is
\begin{equation}\label{PoyntinginCylindricalcloak}
  \langle \vec{S}^\mathrm{c}_{\scriptscriptstyle(\mathrm{P})}\rangle\ =\ \frac{{E_0}^2}{2\eta_0}\left( \frac{f(r)}{r}\cos\theta\hat{e}_r-f'(r)\sin\theta\hat{e}_\theta \right).
\end{equation}
and the time-averaged energy density of electromagnetic fields under the assumption of non-dispersion is
\begin{equation}\label{energydensityinCylindricalcloak}
  \langle W^\mathrm{c}_{\scriptscriptstyle(\mathrm{P})}\rangle\ =\ \frac{\varepsilon_0{E_0}^2f(r)f'(r)}{2r}.
\end{equation}

By calculation, the ray vector and wave vector in the cylindrical
cloak have the identical form in the spherical cloak as shown in Eq.
(\ref{rayvectorinsphericalcloak}) and Eq.
(\ref{wavevectorinsphericalcloak}). Whereas, the variables $r$ and
$\theta$ in the spherical cloak are spherical coordinates, while in
cylindrical cloak they represent cylindrical coordinates.

Thus, following the same calculation as in spherical cloak, we derive the light-ray expression in cylindrical cloak
\begin{equation}\label{lightrayinCylindricalcloak}
  f(r)\sin\theta\ \ =\ \ \mbox{const},\qquad z\ \ =\ \ z_0,
\end{equation}
and the wave-normal ray equation
\begin{equation}\label{wave-normalinCylindricalcloak}
   \exp\left[\int\frac{f(r)}{r^2f'(r)}\dif r\right]\sin\theta\ \ =\ \ \mbox{const},\qquad z\ \ =\ \ z_0.
\end{equation}

\subsection{Examples of four types of cloak}
In this section, we apply the obtained expressions of light-ray and
wave-normal ray to four cloaks constructed by different
transformation functions. Type 1 is the linear transformation
function $f_1(r)=[b/(b-a)](r-a)$ provided in \cite{Pendry2006Sci,
Schurig2006OE}. Type 2 is $f_2(r)=b[(r-a)/(b-a)]^2$, and type 3 is
$f_3(r)=b-b[(b-r)/(b-a)]^2$. Both type 2 and 3 are from
\cite{Luo2008PRB}. Type 4 is
$f_4(r)=(b/2a)[2a-b+\sqrt{b^2-4ab+4ar}]$ provided in
\cite{Cai2007APL}. They all meet the invisibility condition $f(a)=0,\
f(b)=b$ (see Fig. 2).

\begin{figure}[h]
\setlength{\abovecaptionskip}{0pt}
\begin{center}
\includegraphics[width=0.5\columnwidth,clip]{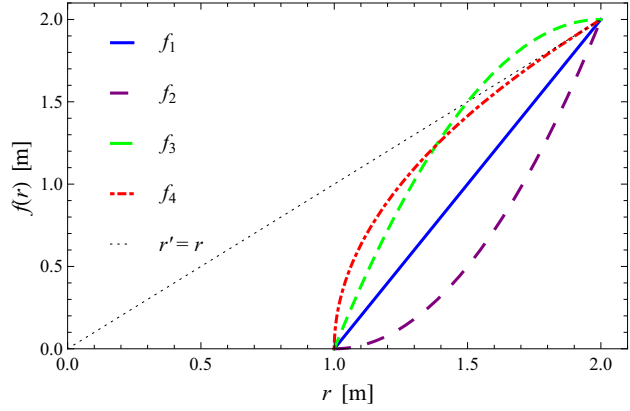}  
\par\end{center}
  \caption{Transformation functions. The inner and outer radiuses of cloak are $a=1\mathrm{m}$ and $b=2\mathrm{m}$ respectively.   }
\end{figure}

We construct four spherical cloaks with the four transformations.
Each raw in Fig. 3 displays a type of cloak corresponding to
$f_1(r)$ to $f_4(r)$ from top to bottom. The first column displays
the components of electric fields
$E_{{\scriptscriptstyle(\mathrm{P})}\langle x \rangle}$ in the $x=0$
plane. The second column displays the light-rays and wave-fronts of
each cloak. The third column shows the wave-normal rays and
wave-fronts which are always perpendicular to each other.

From Fig. 3, some characteristics of each cloak can be concluded. In Fig. 3(b), the light-rays and wave-fronts hold equal spacing respectively, which is due to its linear transformation
\begin{figure}[!htbp]
  \centering
\begin{minipage}[t]{0.355\columnwidth}
  \vspace{0pt}
    \includegraphics[width=5cm]{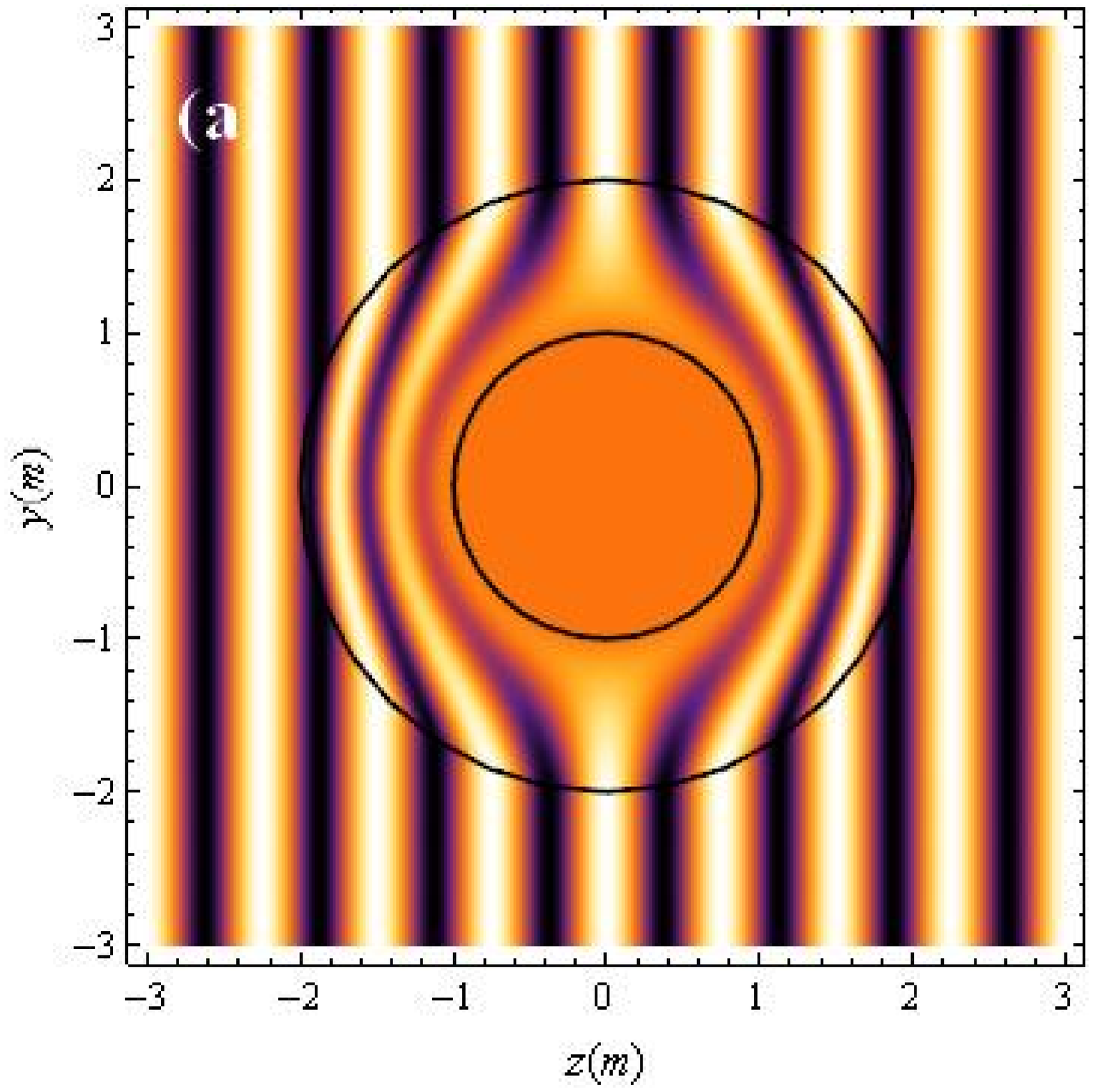}
  \label{fig:subfig:a} 
\end{minipage}
\begin{minipage}[t]{0.315\columnwidth}
 \vspace{0pt}
    \label{fig:subfig:b} 
    \includegraphics[width=4.5cm]{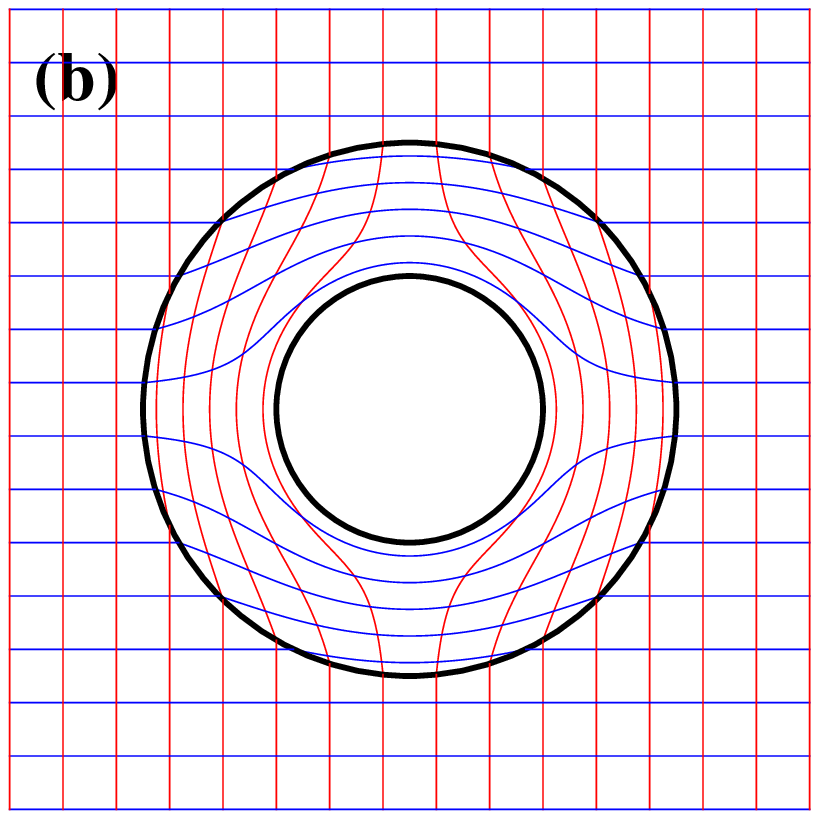}
\end{minipage}
\begin{minipage}[t]{0.315\columnwidth}
 \vspace{0pt}
      \label{fig:subfig:c} 
    \includegraphics[width=4.5cm]{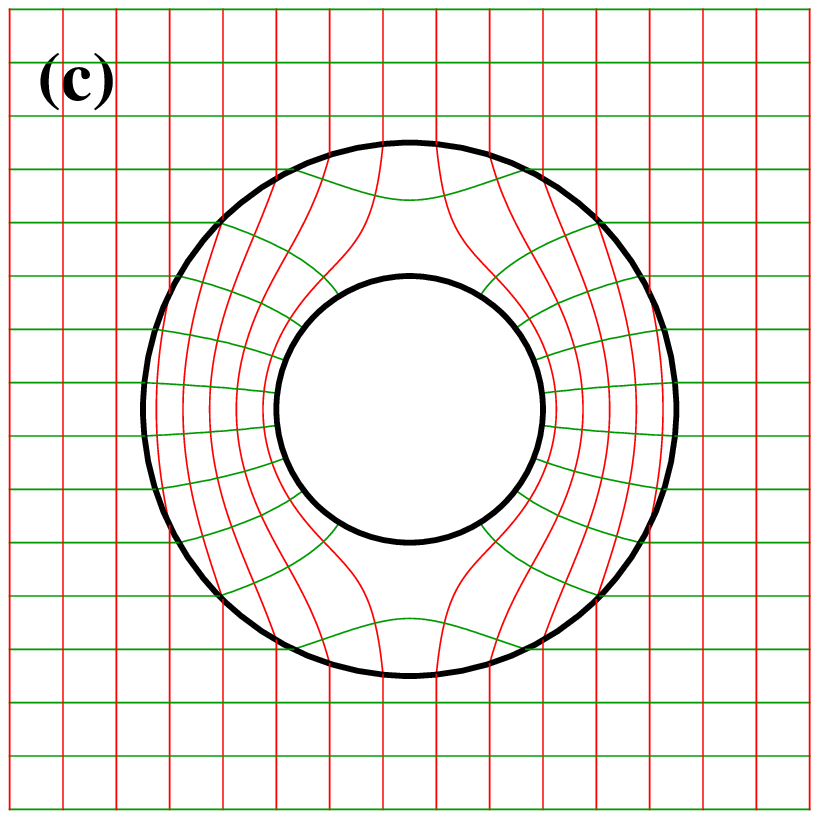}
\end{minipage}
\begin{minipage}[t]{0.355\columnwidth}
  \vspace{0pt}
    \label{fig:subfig:d} 
    \includegraphics[width=5cm]{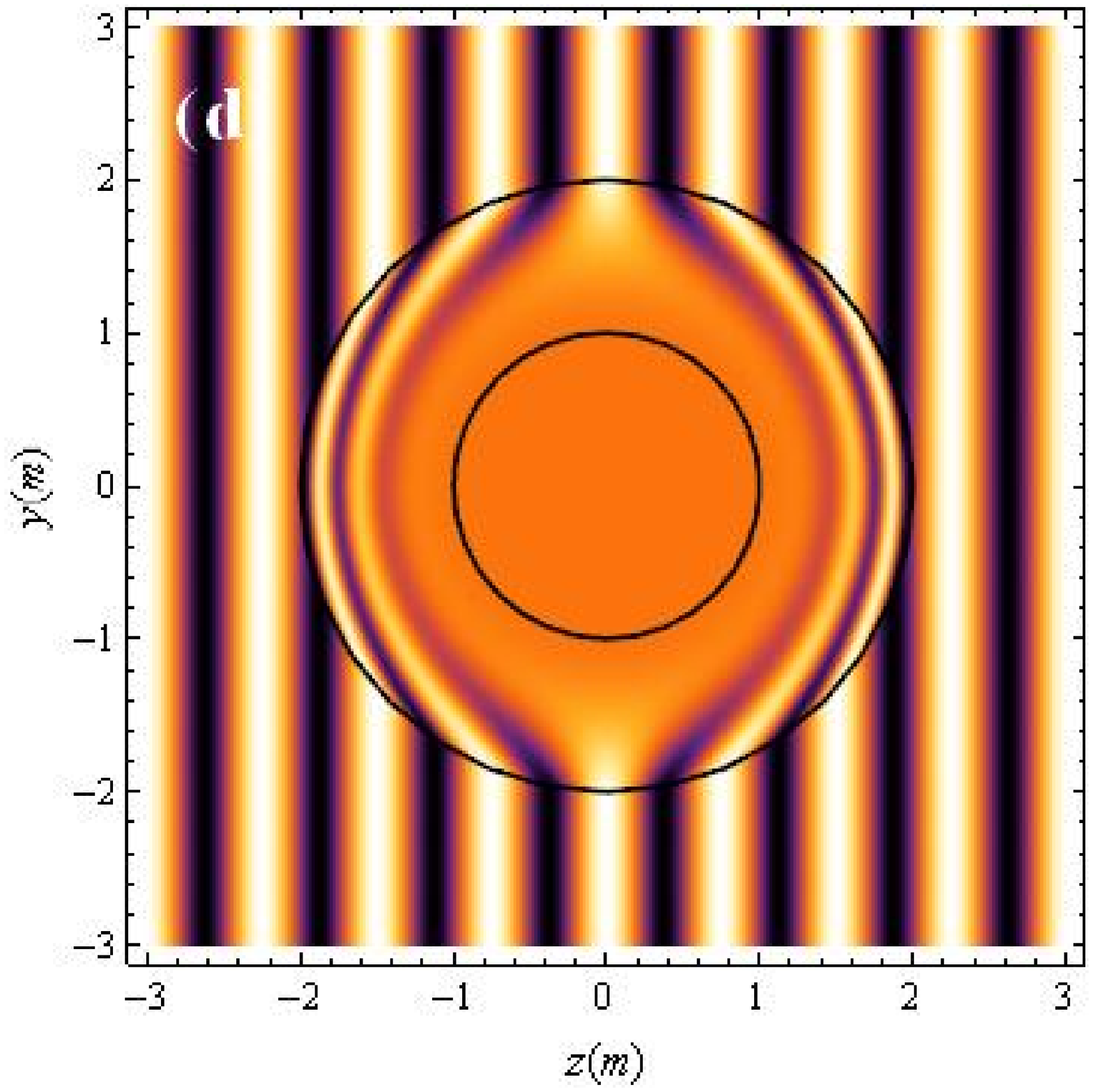}
\end{minipage}
\begin{minipage}[t]{0.315\columnwidth}
 \vspace{0pt}
    \label{fig:subfig:e} 
    \includegraphics[width=4.5cm]{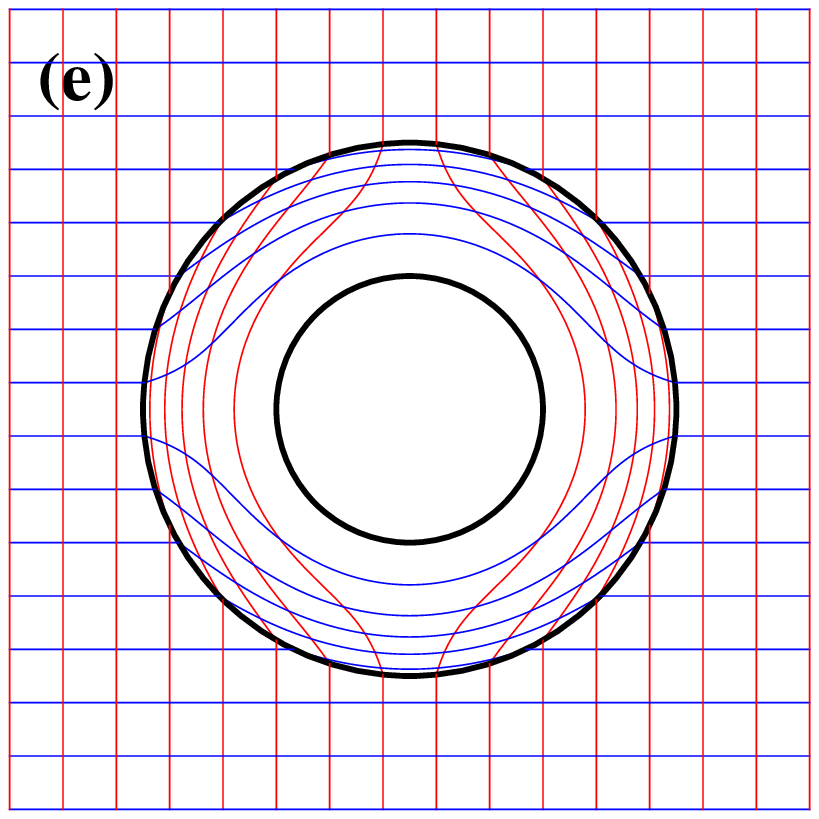}
\end{minipage}
\begin{minipage}[t]{0.315\columnwidth}
 \vspace{0pt}
      \label{fig:subfig:f} 
    \includegraphics[width=4.5cm]{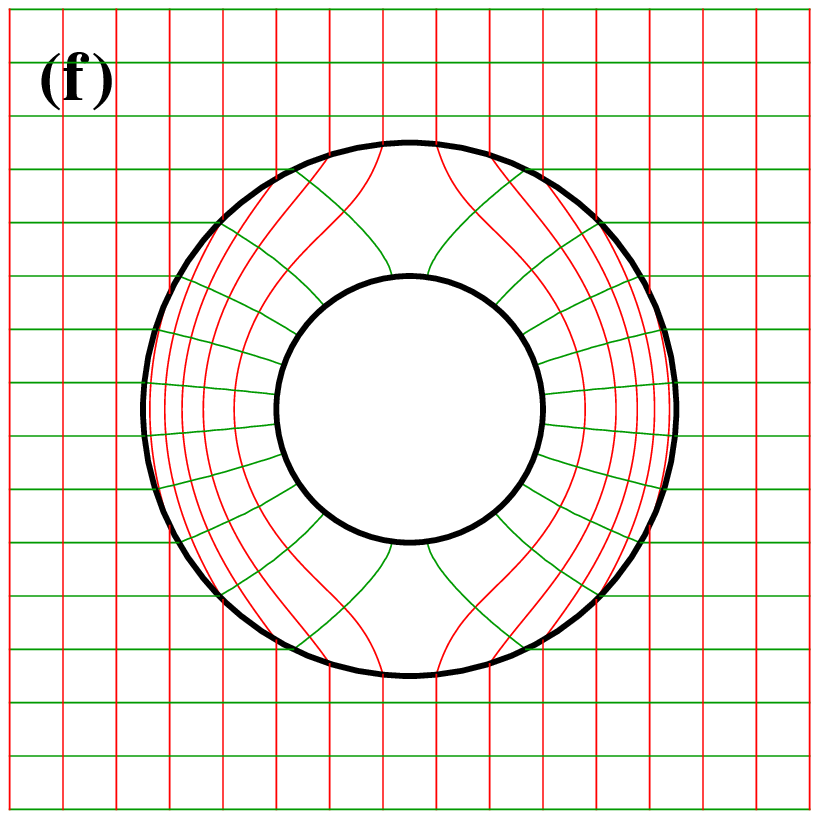}
\end{minipage}
\begin{minipage}[t]{0.355\columnwidth}
  \vspace{0pt}
    \label{fig:subfig:g} 
    \includegraphics[width=5cm]{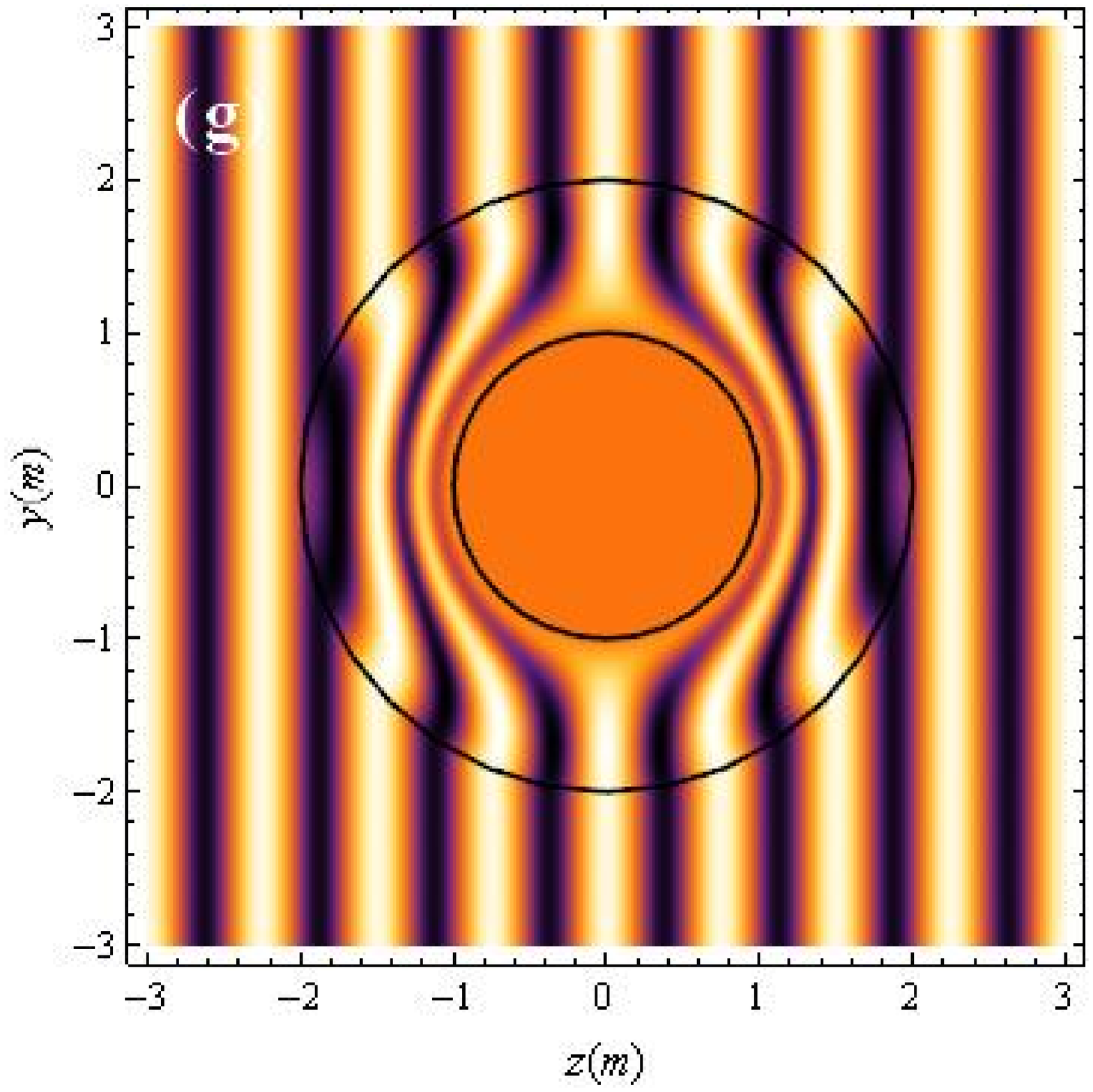}
\end{minipage}
\begin{minipage}[t]{0.315\columnwidth}
 \vspace{0pt}
    \label{fig:subfig:h} 
    \includegraphics[width=4.5cm]{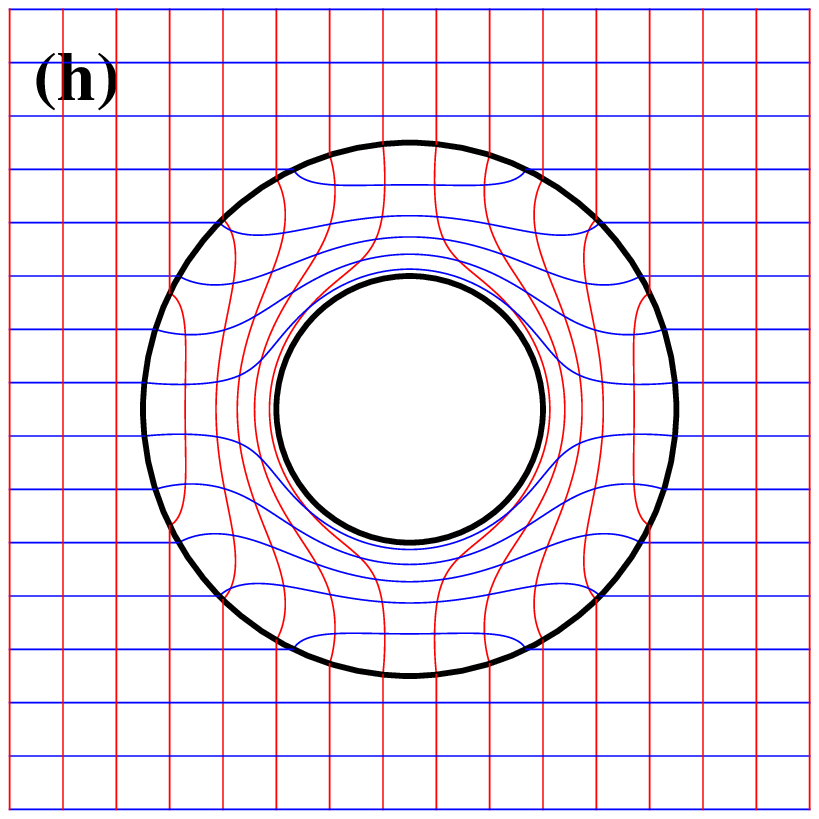}
\end{minipage}
\begin{minipage}[t]{0.315\columnwidth}
 \vspace{0pt}
      \label{fig:subfig:i} 
    \includegraphics[width=4.5cm]{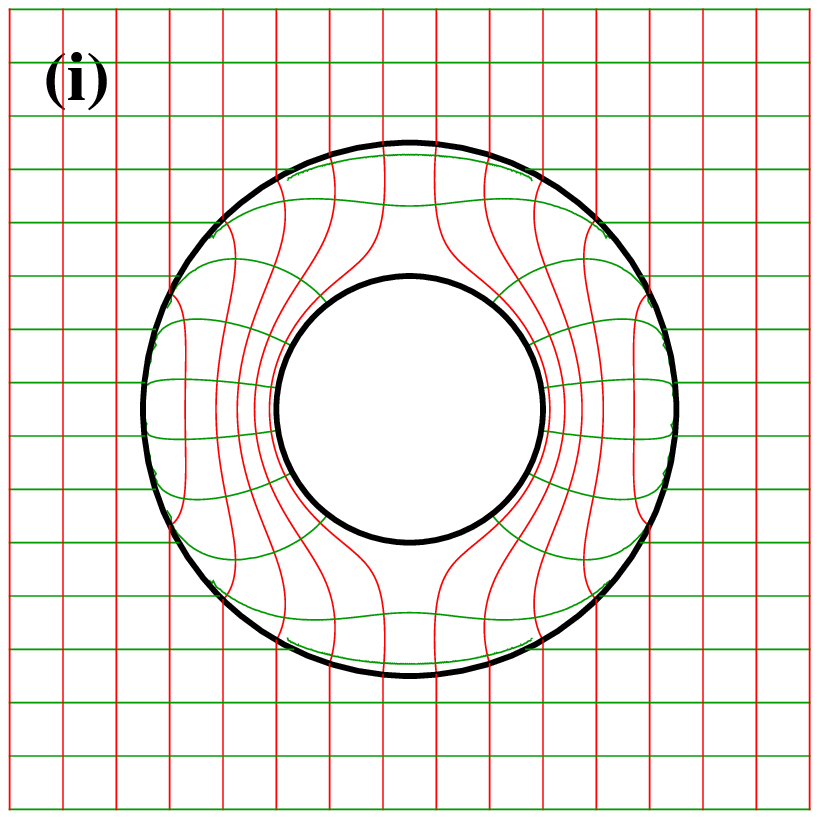}
\end{minipage}
\begin{minipage}[t]{0.355\columnwidth}
  \vspace{0pt}
    \label{fig:subfig:j} 
    \includegraphics[width=5cm]{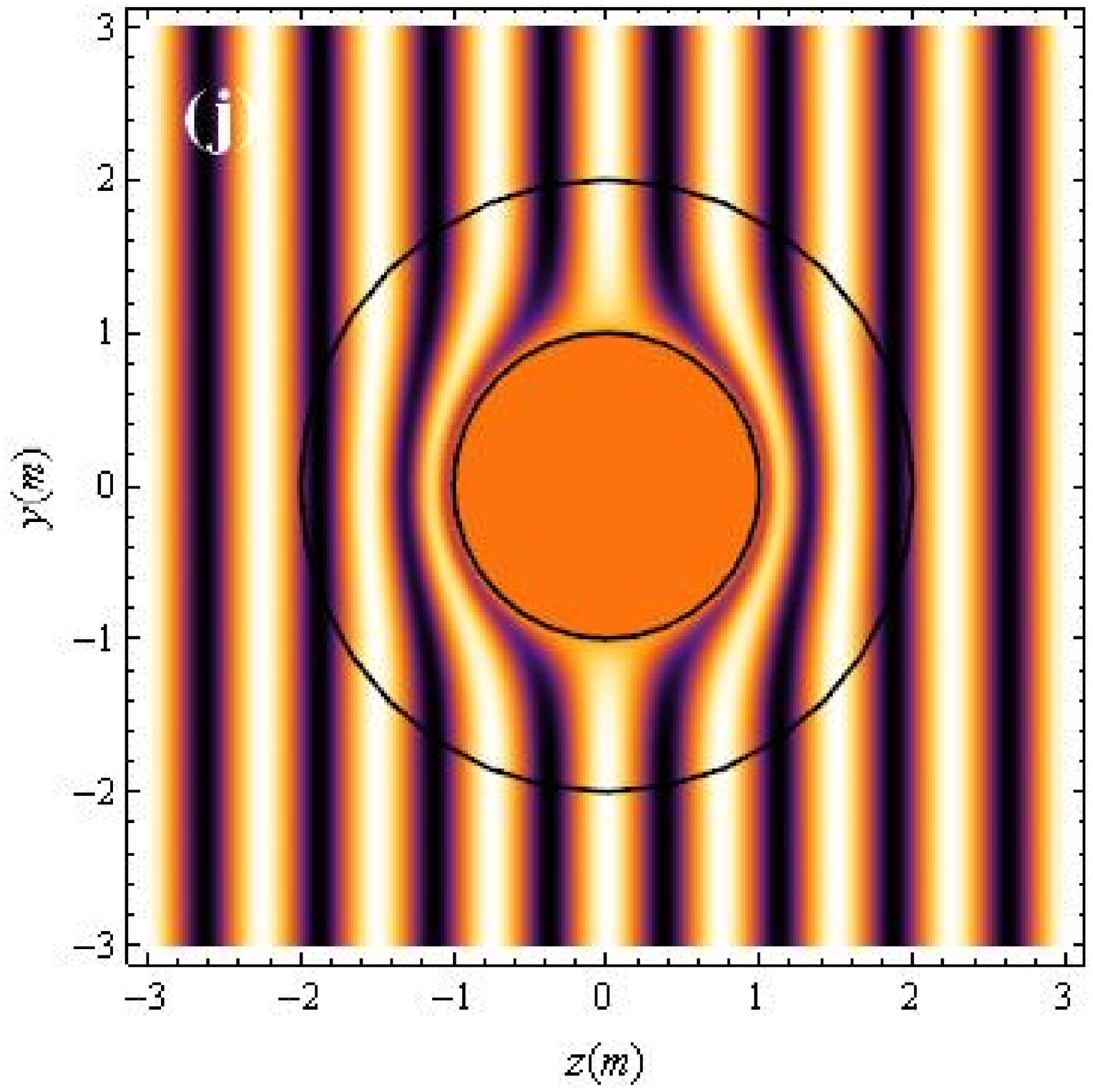}
\end{minipage}
\begin{minipage}[t]{0.315\columnwidth}
 \vspace{0pt}
    \label{fig:subfig:k} 
    \includegraphics[width=4.5cm]{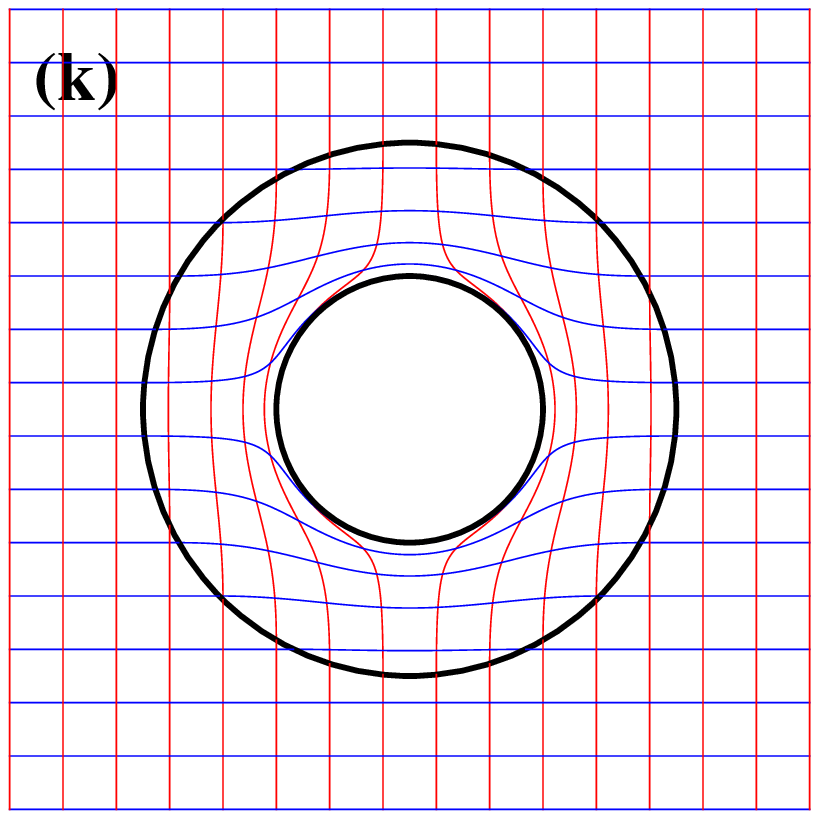}
\end{minipage}
\begin{minipage}[t]{0.315\columnwidth}
 \vspace{0pt}
      \label{fig:subfig:l} 
    \includegraphics[width=4.5cm]{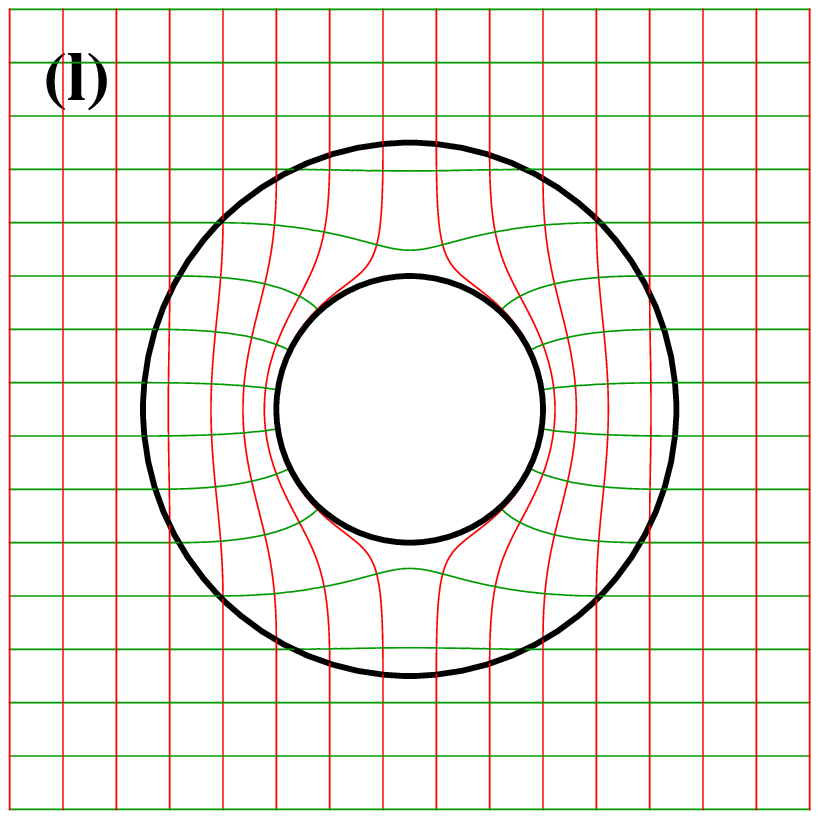}
\end{minipage}
\caption{Lights travel in $x=0$ plane of four space where four types of cloak are placed respectively. Field distributions of $E_{{\scriptscriptstyle(\mathrm{P})}\langle x \rangle}$ of each cloak are shown in (a)(d)(g)(j). In (b)(e)(h)(k), blue lines denote light-rays, red lines denote wave-fronts. In (c)(f)(i)(l), green lines denote wave-normal rays, red lines still denote wave-fronts.   }
\end{figure}
$f_1(r)$. For the second cloak, because $f_2'(r) \rightarrow 0$ as $r \rightarrow a$, the $\theta$ component of energy flow decreases in $r \rightarrow a$ so fast that the energy traveling in cloak mainly concentrates at the outer boundary. Thereby the light-rays mostly compress near the outer boundary shown in Fig. 3(e). For the third cloak, since $f_3'(b)=0$, the light-rays are condensed to inner boundary as shown in Fig. 3(h). In addition, $r=b$ as $k_{{\scriptscriptstyle(\mathrm{P})}r}\equiv0$, that is to say, the circle $r=b$ becomes a closed wave-normal ray, and the wave-normal rays inside cloak are tangent to the outer interface $r=b$, as shown in Fig. 3(i). For the fourth cloak, its $\varepsilon$ and $\mu$ are continuous to vacuum across outer interface due to $f_4'(b)=1$. Therefore, all the light-rays, wave-normal rays and wave-fronts are smooth when they pass through the outer boundary of cloak as shown in Fig. 4(k)(l).

\subsection{Verifying the light-ray equation is the solution of geodesic equations}
In this section, We will verify that the light-ray analytical expression (\ref{lightrayinsphericalcloak}) in a spherical cloak satisfies the geodesic equation in {\bf S} system of virtual space, which is \cite{Niu2009OE}
\begin{equation}\label{geodesicequation}
\frac{\dif k^i}{\dif\lambda}+\Gamma^i_{jl}k^jk^l\ =\ 0,
\end{equation}
where $\Gamma^i_{jl}=\frac{1}{2}\gamma^{im}(\gamma_{mj,l}+\gamma_{ml,j}-\gamma_{jl,m})$ is the Christoffel symbol in 3-D space.
In {\bf S} system, the spatial metric is
\begin{equation}\label{S-ref_metric}
   \gamma_{ij}\ =\ \mathrm{diag} \big( f'(r)^2,\ \ f^2(r),\ \ f^2(r)\sin^2\theta \big).
\end{equation}
Substituting it into Christoffel symbol, we have
\begin{equation}\label{Christoffel}
\left\{
\begin{array}{lll}
  \Gamma^r_{rr}\ =\ \frac{f''(r)}{f'(r)}, & \Gamma^r_{\theta\theta}\ =\ -\frac{f(r)}{f'(r)}, & \hspace{-23pt}\Gamma^r_{\phi\phi}\ =\ -\frac{f(r)}{f'(r)}\sin^2\theta,\\[4pt]
  \Gamma^\theta_{\theta r}\ =\ \frac{f'(r)}{f(r)}, & \Gamma^\theta_{\phi\phi}\ =\ -\sin\theta\cos\theta, & \ \\[4pt]
  \Gamma^\phi_{\phi r}\ =\ \frac{f'(r)}{f(r)}, &  \Gamma^\phi_{\phi\theta}\ =\ \cot\theta, & \ \\[4pt]
  \end{array}
\right.
\end{equation}
and other components are all null. Then, the geodesic equations become
\begin{subequations}\label{geodesicequationinS-ref}
  \begin{flalign}
  & \frac{\dif k^r}{\dif\lambda}+\frac{f''(r)}{f'(r)}k^rk^r-\frac{f(r)}{f'(r)}k^\theta k^\theta-\frac{f(r)}{f'(r)}\sin^\theta k^\phi k^\phi\ =\ 0,\\
  & \frac{\dif k^\theta}{\dif\lambda}+2\frac{f'(r)}{f(r)}k^\theta k^r-\sin\theta\cos\theta k^\phi k^\phi\ =\ 0,\\
  & \frac{\dif k^\phi}{\dif\lambda}+2\frac{f'(r)}{f(r)}k^\phi k^r+\cot\theta k^\theta k^\phi\ =\ 0.
\end{flalign}
\end{subequations}
Tracing the ray path shown in Eq. (\ref{lightrayinsphericalcloak}), the differentials of coordinates satisfy
\begin{displaymath}
   \dif\theta\ \ =\ \ -\frac{f'(r)}{f(r)}\tan\theta\ \dif r,\qquad \dif \phi\ \ =\ \ 0.
\end{displaymath}
Substituting them into the relation between parameter $\lambda$ and element of length $\dif l=(\gamma_{ij}\;\dif x^i\dif x^j)^{1/2}$, which is
$ \dif\lambda\ \ =\ \ -(c\sqrt{-g_{00}}/\omega)  \dif l $ \cite{Landau1995Oxford},
we have
\begin{displaymath}
   \dif\lambda\ \ =\ \ \frac{c f'(r)}{\omega\cos\theta}\  \dif r.
\end{displaymath}
Therefore, the wave vector can be expressed as
\begin{equation}\label{wavevectorinS-ref}
   k^r\ =\ \frac{\dif r}{\dif \lambda}\ =\ k_0\frac{\cos\theta}{f'(r)},\quad\  k^\theta\ =\ \frac{\dif \theta}{\dif \lambda}\ =\ -k_0\frac{\sin\theta}{f(r)},\quad\  k^\phi\ =\ 0.
\end{equation}
This is the wave vector in {\bf S} system, its relation with the
wave vector (\ref{wavevectorinsphericalcloak}) in {\bf P} space will
be discussed in section 4.2. Substituting $k^i$ into Eqs.
(\ref{geodesicequationinS-ref}), then the left sides of the
equations reduce to zero, equal to the right side. As a result, we find
out that Eq. (\ref{lightrayinsphericalcloak}) is a particular
analytical solution of the geodesic equations.

\subsection{Physical interpretation of light-ray equation}
We have verified the light-ray (\ref{lightrayinsphericalcloak})
denotes a geodesic expressed by curved coordinates {\bf S} of
virtual space, thus the physical meaning of the light-ray equation
is quite clear.

Since the virtual space is a flat vacuous space with a zero
curvature, the light-ray trajectory which is also the geodesic in virtual
space is just a straight line. Fig. 4(a) shows the light-ray with a
Cartesian coordinate system ({\bf S$'$} system). Through the inverse
transformation $r'=f(r)$, where $f(a)=0,\ f(b)=b$ inside the domain
$r<b$ in {\bf S$'$} system, we obtain a new system, {\bf S} system.
Through the transformation, the domain $r'<a$ in {\bf S$'$} system
shrinks into origin, that is, the coordinates with component
$r<a$ do not exist in {\bf S} system. Fig. 4(b) shows the result in
{\bf S} system. The coordinate lines, whose tangent vectors are bases, are defined by
\begin{subequations}\label{coordinate lines}
  \begin{flalign}
   \mbox{ Latitude lines :}&\qquad y\ =\ r\sin\theta\ =\ f^{-1}(r')\sin\theta'\ =\ \mbox{const.} \\
   \mbox{ Longitude lines :}&\qquad x\ =\ r\cos\theta\ =\ f^{-1}(r')\cos\theta'\ =\ \mbox{const.}
\end{flalign}
\end{subequations}
\begin{figure}[!htbp]
\setlength{\abovecaptionskip}{0pt}
\begin{center}
    \includegraphics[width=0.6\columnwidth,clip]{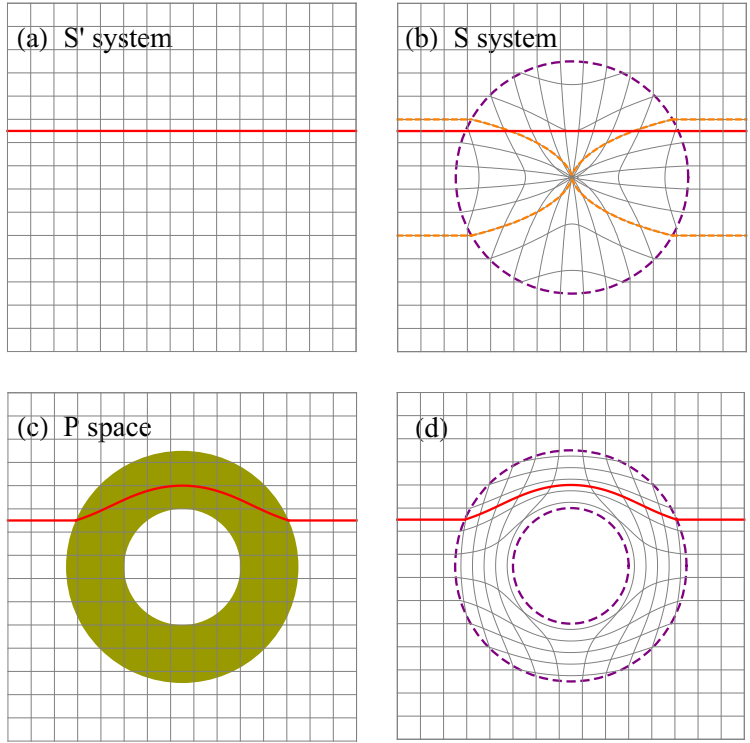}
\par\end{center}
\caption{Interpretation of light-rays. (a) The red line denotes a light-ray propagating in virtual space. The meshes are the coordinate lines of {\bf S$'$} system. (b) The red line is still the light-ray in virtual space. The curved meshes are the coordinate lines of {\bf S} system. Two orange dashed lines are the boundaries inside which latitude lines shrink into the origin point. (c) Light-ray travels in {\bf P} space with Cartesian coordinate meshes. (d) The mapping of lines with the expression of geodesic in {\bf S} system to {\bf P} space.  }
\end{figure}
where $f^{-1}(r')$ is the inverse function of $f(r)$. We observe
that the coordinate lines cave towards the origin, especially the
lines whose $|x|<a$ or $|y|<a$ outside the interface $r=b$ cave
directly into the origin when they pass through the interface.
Therefore, the origin becomes an ambiguous point of coordinates in
{\bf S} system. However, the coordinate transformation doesn't
change the curvature of virtual space, and the light trajectories,
{\it i.e.} geodesics, are still straight with the expression in {\bf
S} system
\begin{equation}\label{light trajectories in S}
   y'\ =\ r'sin\theta'\ =\ f(r)\sin\theta\ =\ \mbox{const.}
\end{equation}
It corresponds to the light-ray equation that we have obtained previously.

The light-ray function maps to {\bf P} space with the same form
$f(r_{\scriptscriptstyle(\mathrm{P})})\sin\theta_{\scriptscriptstyle(\mathrm{P})}=\mbox{const}$.
Since {\bf P} space is also a flat space, the light-ray becomes
convex as shown in Fig. 4(c). Meanwhile, the constitutive parameters both in spherical and cylindrical cloak have singular components with zero or infinite value at inner surface for $f(a)=0$. In other words, the geometrical singularity of coordinates turns into the singularity of material, when mapping from {\bf S} system to {\bf P} space.

We use an analogy to describe the process. The virtual space can be
compared to a piece of flat elastic cloth with straight meshes on
it. The meshes look like {\bf S$'$} system. If we fix the
boundary $r=b$, and drag the whole region $r<a$ into the center
point, the meshes on the cloth are curved as the coordinate lines of
{\bf S} system. Then, we draw a straight line on the cloth just as
the red line in Fig 4(b) which plays the role of light-ray in {\bf
S}. After that, we loose the piece of cloth held into
the center, and let it return flat. Then the meshes also
returns to straightness, but the red line we have drawn becomes
convex exactly as the ray propagating in {\bf P} space.

Now we should clarify the meaning of Fig. 4(d). Sometimes the meshes in this figure are  interpreted as the coordinate of {\bf S} system \cite{Schurig2006OE}. This interpretation gives rise to an illusion that {\bf S} system is a real curved space, as well as the light trajectories. However, the meshes in Fig. 4(d) are neither coordinate lines nor geodesics no matter in {\bf S} system or in {\bf P} space. The only reasonable interpretation is that they are the curves in {\bf P} space directly mapping from the straight lines in {\bf S} system, and can describe the light trajectories in {\bf P} space \cite{Leonhardt2009PO, Chen2010NatMat}.

\section{Duality principle in anisotropic media}
In virtual space,  light propagates in straight lines, and the directions of wave vectors and ray vectors are always identical, however when mapping to physical space, they are not parallel to
each other any more. So what exactly causes the split? Why does the light-ray instead of not the wave-normal ray in {\bf P} space inherit its expression in {\bf S} system? In this section, we will try to answer these questions by using duality principle.

In charge free anisotropic media, electromagnetic fields have the following duality rule \cite{Born1999Cambridge}
\begin{displaymath}
\mbox{Duality rule}
\left\langle \begin{array}{cccccccc}
                   \quad\vec{D}  &\quad  \vec{E}&\quad  \vec{B}  &\quad  \vec{H}  &\quad  \matr{\varepsilon}  &\quad  \matr{\mu}  &\quad  \vec{k}/\omega  &\quad  v_{p}  \\
                   \\
                \quad\vec{E}  &\quad  \vec{D}&\quad  \vec{H}  &\quad  \vec{B}  &\quad  {\matr{\varepsilon}}^{-1}  &\quad  {\matr{\mu}}^{-1}  &\quad  \vec{s}\omega  &\quad  1/v_{ray}  \\
\end{array}
\right.
\end{displaymath}
where $v_{p}=\omega/k$ is the phase speed. 
In terms of the above systematic interchange relations, the either side satisfies a set of formulas for plane waves 
\cite{Chen1985McGraw}:

\begin{equation}\label{duality rule}
\begin{aligned}
&\vec{D} \ = \ \matr{\varepsilon}\cdot\vec{E} \ = \ -\frac{1}{\omega}\vec{k}\times\vec{H}& &\vec{E} \ = \ \matr{\varepsilon}^{-1}\cdot\vec{D} \ = \ -\omega\vec{s}\times\vec{B}\\[5pt]
&\vec{B} \ = \ \matr{\mu}\cdot\vec{H} \ = \ \frac{1}{\omega}\vec{k}\times\vec{E}& \ \ \Leftarrow\ \mbox{duality} \ \Rightarrow \ \ \ \ &\vec{H} \ = \ \matr{\mu}^{-1}\cdot\vec{B} \ = \ \omega\vec{s}\times\vec{D}\\[3pt]
&\vec{k}/\omega \ = \ \hat{k}/v_p \ = \  \frac{\vec{D}\times\vec{B}}{\vec{E}\cdot\vec{D}}& &\vec{s}\omega \ = \ v_{ray}\hat{s} \ = \  \frac{\vec{E}\times\vec{H}}{\vec{D}\cdot\vec{E}}\\
\end{aligned}
\end{equation}
According to the left side of the formulas, the wave equation is founded as
\begin{equation}\label{wave equation}
   \vec{k}\times\left[\ \matr{\mu}^{-1}\cdot\big(\ \vec{k}\times\vec{E}\ \big)\ \right]+\omega^2\ \matr{\varepsilon}\cdot\vec{E}\ =\ 0.
\end{equation}
Concerning the special case of impedance matched material $ \matr{\varepsilon}/\varepsilon_0=\matr{\mu}/\mu_0=\matr{n} $, the dispersion relation, which are also called wave-vector eikonal equation, is (derived in detail in Appendix C)\cite{Schurig2006OE}
\begin{equation}\label{dispersion relation}
   \bar{n}^{ij}k_ik_j-\gamma\left(\frac{\omega}{c}\right)^2\det\left( \bar{n}^{ij} \right)\ =\ \bar{n}_{ij}k^ik^j-\frac{1}{\gamma}\left(\frac{\omega}{c}\right)^2\det\left( \bar{n}_{ij} \right)\ =\ 0.
\end{equation}
Here, there are two ambiguities to be clarify. First, the expression given in \cite{Schurig2006OE} is based on Cartesian coordinates, so $\gamma$
reduces to unit, while the expression here is valid in any curved coordinates. Second, $\bar{n}^{ij}$ and $\bar{n}_{ij}$ presented above express the components in holonomic coordinate bases (Landau's definition), however, the dispersion relations expressed either in Minkowski's definition or in anholonomic unit basis are a little different from Eq. (\ref{dispersion relation}) (see Appendix C). In \cite{Schurig2006OE}, the three kinds of expression are congruent with each other in Cartesian coordinates.

Using the duality principle, we can easily obtain the dual form of wave equation
\begin{equation}\label{dual wave equation}
   \vec{s}\times\left[\ \matr{\mu}\cdot\big(\ \vec{s}\times\vec{D}\ \big)\ \right]+\frac{1}{\omega^2}\ \matr{\varepsilon}^{-1}\cdot\vec{D}\ =\ 0.
\end{equation}
Thus, the ``ray-vector eikonal equation'' is
\begin{equation}\label{dual dispersion relation}
   {(\bar{n}^{\scriptscriptstyle-1})}^{ij}s_is_j-\gamma\left(\frac{c}{\omega}\right)^2\det\left[ (\bar{n}^{\scriptscriptstyle-1})^{ij} \right]\ =\ {(\bar{n}^{\scriptscriptstyle-1})}_{ij}s^is^j-\frac{1}{\gamma}\left(\frac{c}{\omega}\right)^2\det\left[ (\bar{n}^{\scriptscriptstyle-1})_{ij} \right]\ =\ 0,
\end{equation}
where $(\bar{n}^{\scriptscriptstyle-1})^{ij}$  is the inverse matrix of $\bar{n}_{ij}$, and it is easy to verify that its corresponding covariant components $(\bar{n}^{\scriptscriptstyle-1})_{ij}$ are also the inverse matrix of $\bar{n}^{ij}$.

\subsection{Eikonal equation in curved coordinate system {\textbf S}}
We first consider the relation between $k^i$ and $s^i$ in curved coordinate system {\bf S}. In {\bf S} system, the relative constitutive parameter is $ \bar{n}^{ij}=(\bar{n}^{\scriptscriptstyle-1})^{ij}=\gamma^{ij}$. Substituting it into Eq. (\ref{dispersion relation}) and Eq. (\ref{dual dispersion relation}), we get the wave-vector eikonal equation in {\bf S} system
\begin{equation}\label{dispersion relation in S}
   \gamma^{ij}k_ik_j-\left(\frac{\omega}{c}\right)^2\ =\ \gamma_{ij}k^ik^j-\left(\frac{\omega}{c}\right)^2\ =\ 0,
\end{equation}
and the ray-vector eikonal equation
\begin{equation}\label{dual dispersion relation in S}
   \gamma^{ij}s_is_j-\left(\frac{c}{\omega}\right)^2\ =\ \gamma_{ij}s^is^j-\left(\frac{c}{\omega}\right)^2\ =\ 0.
\end{equation}
The two equations indicate that $|\vec{k}|=k_0$ and $|\vec{s}|=1/k_0$. By considering the general relation $\vec{k}\cdot\vec{s}=\gamma_{ij}k^is^j=1$ \cite{Chen1985McGraw}, we can conclude that the directions of $\vec{k}$ and $\vec{s}$ completely coincide, and they have the relation
$ k_i={k_0}^2s_i$.
This result is actually not marvelous, since the electromagnetic wave is merely propagating in a vacuous space, despite the space is expressed in a curvilinear coordinate system with non-unit metric $\gamma_{ij}$. 

\subsection{Eikonal equation in physical space {\textbf P}}
 In  physical space, the media of the cloak are inhomogeneous and waves traveling in it are no longer plane waves. Nevertheless, we can verify the fields obtained both in the spherical and cylindrical cloak still satisfy Eqs. (\ref{duality rule}, \ref{wave equation}, \ref{dual wave equation}). The relative constitutive parameter of the cloak is $ {\bar{n}_{\scriptscriptstyle(\mathrm{P})}}^{ij}=\sqrt{\gamma/\gamma_{\scriptscriptstyle(\mathrm{P})}}\ \gamma^{ij}$, and its inverse is $({\bar{n}^{\scriptscriptstyle-1}_{\scriptscriptstyle(\mathrm{P})}})_{ij}=\sqrt{\gamma_{\scriptscriptstyle(\mathrm{P})}/\gamma}\ \gamma_{ij}$. Substituting them into Eq. (\ref{dispersion relation}) and Eq. (\ref{dual dispersion relation}), we obtain the wave-vector eikonal equation in {\bf P} space
\begin{equation}\label{dispersion relation in P}
   \gamma^{ij}{k_{\scriptscriptstyle(\mathrm{P})}}_i{k_{\scriptscriptstyle(\mathrm{P})}}_j-\left(\frac{\omega}{c}\right)^2\ =\ \big(\gamma_{{\scriptscriptstyle(\mathrm{P})}ik}\gamma_{{\scriptscriptstyle(\mathrm{P})}il} \gamma^{kl} \big){k_{\scriptscriptstyle(\mathrm{P})}} ^i{k_{\scriptscriptstyle(\mathrm{P})}}^j-\left(\frac{\omega}{c}\right)^2\ =\ 0,
\end{equation}
and the ray-vector eikonal equation in {\bf P} space
\begin{equation}\label{dual dispersion relation in P}
   \gamma_{ij}{s_{\scriptscriptstyle(\mathrm{P})}}^i{s_{\scriptscriptstyle(\mathrm{P})}}^j-\left(\frac{c}{\omega}\right)^2\ =\ \big({\gamma_{{\scriptscriptstyle(\mathrm{P})}}}^{ik}{\gamma_{{\scriptscriptstyle(\mathrm{P})}}}^{jl} \gamma_{kl} \big){s_{\scriptscriptstyle(\mathrm{P})}} _i{s_{\scriptscriptstyle(\mathrm{P})}}_j-\left(\frac{c}{\omega}\right)^2\ =\ 0.
\end{equation}

Comparison of Eqs. (\ref{dispersion relation in P}, \ref{dual dispersion relation in P}) and Eqs.  (\ref{dispersion relation in S}, \ref{dual dispersion relation in S}) manifests that the wave-vector eikonal equations expressed by covariant components $k_i$ have the same form in {\bf P} space and {\bf S} system. However if we write the wave-vector eikonal equations with contravariant components $k^i$, the expressions in {\bf P} and in {\bf S} are different. Similarly, the expressions with contravariant components $s^i$ have the same form in the two systems, yet the expressions with covariant components $s_i$ are different. Considering the relation  $\vec{k}_{\scriptscriptstyle(\mathrm{P})}\cdot\vec{s}_{\scriptscriptstyle(\mathrm{P})}=1$ simultaneously, we can get
\begin{equation}\label{wave-vector and ray-vector in P}
   {k_{\scriptscriptstyle(\mathrm{P})}}_i\ \ =\ \ {k_0}^2\; \gamma_{ij}{s_{\scriptscriptstyle(\mathrm{P})}}^i,\qquad
   {s_{\scriptscriptstyle(\mathrm{P})}}^i\ \ =\ \ {k_0}^{-2}\; \gamma^{ij}{k_{\scriptscriptstyle(\mathrm{P})}}_i.
\end{equation}
This relation is easily validated according to Eq. (\ref{rayvectorinsphericalcloak}) and (\ref{wavevectorinsphericalcloak}). It is also valid in {\bf S} system which means $\vec{k}$ is parallel to $\vec{s}$, however this does not make sense in invisibility cloak.

These results also can be verified with canonical equations given by \cite{Schurig2006OE}. If we define Hamiltonian by wave-vector eikonal:
\begin{equation}\label{Hamiltonian}
   H\ =\ \gamma^{ij}{k_{\scriptscriptstyle(\mathrm{P})}}_i{k_{\scriptscriptstyle(\mathrm{P})}}_j-\left(\frac{\omega}{c}\right)^2,
\end{equation}
and according to the first equation of canonical equations
\begin{equation}\label{canonical equations}
   \frac{\dif x_{\scriptscriptstyle(\mathrm{P})}^i}{\dif \tau}\ =\ \frac{\partial H}{\partial k_{{\scriptscriptstyle(\mathrm{P})}i}},\qquad  \frac{\dif k_{{\scriptscriptstyle(\mathrm{P})}i}}{\dif \tau}\ =\ -\frac{\partial H}{\partial x_{\scriptscriptstyle(\mathrm{P})}^i},
\end{equation}
where $\tau$ is the parameter of canonical equations, we obtain the tangent vector of light-ray
 $\dif x_{\scriptscriptstyle(\mathrm{P})}^i/{\dif \tau} = 2\gamma^{ij}k_{{\scriptscriptstyle(\mathrm{P})}j}.$
It reveals the tangent of light-ray is exactly towards the direction of ${s_{\scriptscriptstyle(\mathrm{P})}}^i$ given in Eq. (\ref{wave-vector and ray-vector in P}). If we define another Hamiltonian by ray-vector eikonal:
\begin{equation}\label{Hamiltonian2}
   \tilde{H}\ =\ \big({\gamma_{{\scriptscriptstyle(\mathrm{P})}}}^{ik}{\gamma_{{\scriptscriptstyle(\mathrm{P})}}}^{jl} \gamma_{kl} \big){s_{\scriptscriptstyle(\mathrm{P})}} _i{s_{\scriptscriptstyle(\mathrm{P})}}_j-\left(\frac{c}{\omega}\right)^2,
\end{equation}
and modify the canonical equations into
\begin{equation}\label{canonical equations 2}
   \frac{\dif x_{\scriptscriptstyle(\mathrm{P})}^i}{\dif \tau}\ =\ \frac{\partial \tilde{H}}{\partial s_{{\scriptscriptstyle(\mathrm{P})}i}},\qquad  \frac{\dif s_{{\scriptscriptstyle(\mathrm{P})}i}}{\dif \tau}\ =\ -\frac{\partial \tilde{H}}{\partial x_{\scriptscriptstyle(\mathrm{P})}^i},
\end{equation}
we would have
\begin{equation}
   \frac{\dif x_{\scriptscriptstyle(\mathrm{P})}^i}{\dif \tau}\ =\ 2{\gamma_{{\scriptscriptstyle(\mathrm{P})}}}^{ik}{\gamma_{{\scriptscriptstyle(\mathrm{P})}}}^{jl} \gamma_{kl}{s_{\scriptscriptstyle(\mathrm{P})}}_j\ =\ 2{k_0}^{-2}k_{\scriptscriptstyle(\mathrm{P})}^i.
\end{equation}
As a result, the modified canonical equations depict the trace of wave-normal rays.

Actually, if we express $k_i$ and $s^i$ with $D^i,\ B^i,\ E_i,\ H_i$ in Minkowski's definition, we have
\begin{equation}\label{wave-vector and ray-vector in Minkowski definition}
   k_i\ =\ \omega\frac{e_{ijk}D^jB^k}{E_iD^i},\qquad s^i\ =\ \frac{1}{\omega}\frac{e^{ijk}D_jB_k}{E_iD^i}.
\end{equation}
The expressions 
apply to both {\bf S} system and {\bf P} space. According to the relations in Table \ref{Table: Corresponding relation}, we can get two new corresponding relations between {\bf S} and {\bf P}:
\begin{displaymath}
   k_i\ \Leftrightarrow\ k_{{\scriptscriptstyle(\mathrm{P})}i},\qquad s^i\ \Leftrightarrow\ s_{{\scriptscriptstyle(\mathrm{P})}}^i.
\end{displaymath}
We note that $k^i$ and $k_{{\scriptscriptstyle(\mathrm{P})}}^i$ do
not correspond to each other, and neither do $s_i$ and
$s_{{\scriptscriptstyle(\mathrm{P})}i}$. In other words, covariant
components of wave vector $k_i$ in {\bf S} system map to wave vector
$k_{{\scriptscriptstyle(\mathrm{P})}i}$ in {\bf P} space, yet
contravariant components of wave vector $k^i\ (={k_0}^2s^i)$ map to
ray vector $s_{{\scriptscriptstyle(\mathrm{P})}}^i$ in  {\bf P}
space. Although $k_i,\ s^i$ denote a same orientation in {\bf S}
system, $k_{{\scriptscriptstyle(\mathrm{P})}i}, and \
s_{{\scriptscriptstyle(\mathrm{P})}}^i$ denote two different
orientations in {\bf P} space. In addition, the contravariant
components $s^i=\dif x^i/\dif \lambda$ directly represent the
directional change of coordinates $\dif x^i$, therefore light-ray
equation in {\bf P} space  inherits itself's expression in {\bf S}.
By contrast, $k_i$ can not directly evince the change of $\dif x^i$,
so the expression of wave-normal ray takes a new form in {\bf P}
space.

According to the dispersion relation (\ref{dispersion relation in
P}), we can obtain the group velocity in the cloak through applying
the formula $\vec{v}_{g}=\nabla_{\scriptscriptstyle\vec{k}}\omega$,
where $\nabla_{\scriptscriptstyle\vec{k}}$ is the divergence
operator in $\vec{k}$--space. Supposing the medium is
non-dispersive, the group velocity takes the identical expression of
ray velocity (\ref{rayvelocityinsphericalcloak}). It means the group
velocity and ray velocity are congruent in the cloak.

As commented in section 3.1, the contradiction of super velocity of
light takes into account the existence of dispersion. However, the
dispersive processes of $\varepsilon$ and $\mu$ are different, so
the case of impedance matching established in a particular frequency
would hardly retain when frequency changes. Therefore, the
dispersion relation of impedance matched material (\ref{dispersion
relation}) is unavailable when we use
$\vec{v}_{g}=\nabla_{\scriptscriptstyle\vec{k}}\omega$ to calculate
the group velocity in dispersive case, because Eq. (\ref{dispersion
relation}) can not display the differences between the derivative of
$\varepsilon$ and $\mu$ with respect to $\omega$.

\section{Design of the invisibility cloak with rays satisfying harmonic function }
In Ref. \cite{Leonhardt2006Sci}, Leonhardt applied \font\fontWCA=wncyr10 {\fontWCA  Zhukovski\ae}
 function $\zeta(z)=z+1/z$, where $z=r\ \me^{\mi \theta}$, to a conformal mapping, and obtained an inhomogeneous but isotropic medium which can realize an invisibility device in geometric  limit. In the media, the families of light-rays and of wave-fronts are the imaginary and real part of $\zeta(z)$ respectively. In addition, they are both harmonic functions and orthogonal to each other.

 Using the expression of light-ray (\ref{lightrayinCylindricalcloak}), we can design an invisibility cloak by means of transformation-optics method, which  gives the same light trajectories as in Leonhardt's device. However, we start our work with a more general target to find the general form of transformation function $f(r)$ which causes the light-rays in cylindrical cloak to satisfy harmonic function. Substitution of light-ray (\ref{lightrayinCylindricalcloak}) into Laplace's equation yields
\begin{equation}\label{Laplace equation}
  \nabla_{\scriptscriptstyle 2-D}\ \left[\ f(r)\sin\theta\ \right]\ =\ 0.
\end{equation}
Then, it reduces to Euler's differential equation, $
r^2f''(r)+rf'(r)-f(r)=0 $, therefore the general solution is $ f(r)
= C_1(r^2-C_2)/r$. The function has a zero point $
r={C_2}^{\scriptscriptstyle1/2}>0 $, which is the most important
condition to construct an invisibility cloak. If we set $
C_1=b^2/(b^2-a^2) $ and $C_2=a^2$, the transformation function
becomes
\begin{equation}\label{transformation for Laplace}
  f(r)\ =\ \frac{b^2}{b^2-a^2}\;\frac{r^2-a^2}{r}.
\end{equation}

\begin{figure}[t]
\setlength{\abovecaptionskip}{0pt}
\begin{center}
    \includegraphics[width=0.8\columnwidth,clip]{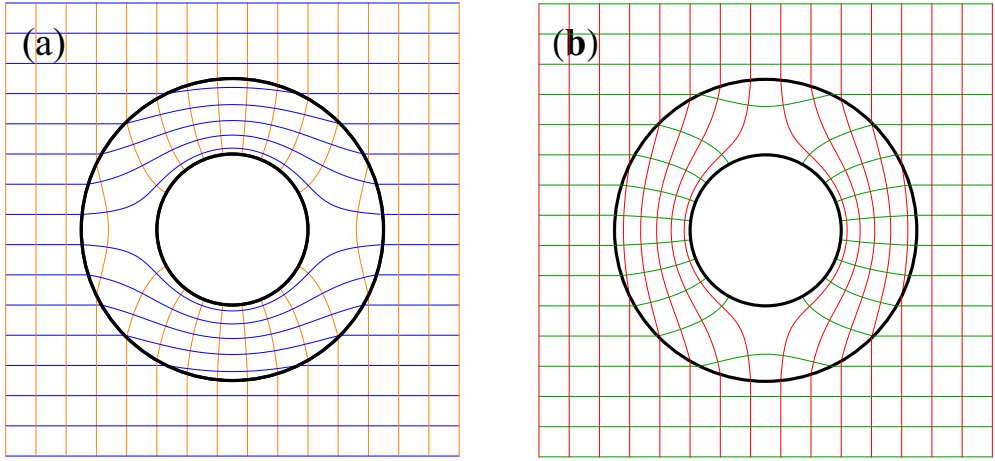}
\par\end{center}
\caption{Cloak with rays satisfying harmonic equation. (a) Blue lines denote light-rays, orange lines denote the surfaces orthogonal to light-rays. In cloak layer, they are painted by Eq. (\ref{analytical function1}). (b) Green lines denote wave-normal rays, red lines denote wave-fronts. In cloak layer, they are painted by Eq. (\ref{analytical function2}).  }
\end{figure}
\setcounter{subfigure}{0}

The result evinces that each solution corresponds  to an
invisibility cloak with a spacial pair of inner and outer radii. In
the process of derivation, we didn't command $f(a)=0, f(b)=b$,
however the result indicates that the form of harmonic functions
$V(r,\theta)=f(r)\sin\theta$ is unique, and it always can realize
invisibility cloak. For convenience, we omit the coefficient of
$f(r)$ in the following derivation which doesn't influence the shape
of rays.

$V=\mbox{const}$ represents the family of light-rays in the cloak.
Let $V$ be the imaginary part of an analytical function $F(z)$.
Using Cauchy-Riemann condition, we obtain the real part
$U(r,\theta)=[(r^2+a^2)/r]\cos\theta$.The $U=\mbox{const}$ stands
for a family of curved surfaces, which is orthogonal to the energy
flows. The analytical function $F(z)$  is
\begin{equation}\label{analytical function1}
  F(z)\ =\ z\ +\ \frac{a^2}{z}.
\end{equation}
If we simply set $z'=z/a$, then $F(z)=a\zeta(z')$, where $\zeta(z)=z+1/z$ is precisely
\font\fontWCA=wncyr10 {\fontWCA  Zhukovski\ae}
 function. It is exactly the conformal mapping which Leonhardt used to design the invisibility devices \cite{Leonhardt2006Sci}. The complex analytical function is a compact way to describe the ray path by the meaning of real and imaginary part explained above, as is shown in Fig. 5(a).

Regarding to the transformation (\ref{transformation for Laplace}),
the eikonal $\psi=k_0f(r)\cos\theta$ is a harmonic function as well.
So we also can compress the wave-fronts and wave-normal rays into
another analytical function
\begin{equation}\label{analytical function2}
  G(z)\ =\ z\ -\ \frac{a^2}{z}.
\end{equation}
The real part represents the family of wave fronts, and the
imaginary part represents the family of wave-normal rays exhibited
in Fig. 5(b).

Owing to the energy conservation, the time-averaged Poynting vector
$\langle\vec{S}^\mathrm{c}_{\scriptscriptstyle(\mathrm{P})}\rangle$
is divergence-free. And the wave vector must be irrotational, since
$\vec{k}_{\scriptscriptstyle(\mathrm{P})}=\nabla\psi$. However, if and only
if the transformation satisfies Eq. (\ref{transformation for
Laplace}), we can prove both
$\langle\vec{S}^\mathrm{c}_{\scriptscriptstyle(\mathrm{P})}\rangle$
and $\vec{k}_{\scriptscriptstyle(\mathrm{P})}$ are not only
divergence-free but also irrotational. This conclusion is
consistent to the fact that both light-rays and wave-normal rays
satisfy Laplace's equation for the particular transformation
function.

\section{Conclusion}
In conclusion, we have meticulously compared the behaviors of electromagnetic fields in {\bf P} space with in {\bf S} system of virtual space, and provided the corresponding relations with electromagnetic quantities and boundary conditions between the two systems. Then, we obtain the analytical expression of light-rays and wave-normal rays, and give their physical interpretation. Furthermore, we verify the expression of light-rays we obtained is exactly a particular solution of geodesic equation in {\bf S} system. In addition, using duality principle, we show that the covariant and contravariant components of wave vector in {\bf S} system map to wave vector and ray vector respectively in {\bf P} space, and it causes the split of $k_{{\scriptscriptstyle(\mathrm{P})}i}$ and $s_{{\scriptscriptstyle(\mathrm{P})}}^i$  in {\bf P} space. Finally, we find out the general form of transformation function which turns the light-ray to harmonic, and therefore we can construct a special cloak in which $\langle\vec{S}^\mathrm{c}_{\scriptscriptstyle(\mathrm{P})}\rangle$ and $\vec{k}_{\scriptscriptstyle(\mathrm{P})}$ are both divergence-free and curl-free.

\section*{Acknowledgment}
We thank K. Niu for the help with composing the article. Meanwhile, we also thank C. Song, Y.-B. Hu, L. Fang, J.-N. Zhang, and J. Dai for the helpful discussion. This work was supported in part by NSF of China (Grants No. 11075077), and by SRFDP (Grants No. 200800550015).

\newcounter{appendix}
\numberwithin{appendix}{section}
\renewcommand\theappendix{\Alph{appendix}}
\renewcommand\theequation{\theappendix.\arabic{equation}}%

\section*{Appendix A:\quad Different way to define electromagnetic fields in curved spacetime}
\stepcounter{appendix}
\setcounter{equation}{0}

In Minkowski spacetime, 3-D macroscopic Maxwell's equations in a Lorentz frame are
\begin{subequations}\label{Maxwelleqs3Dflat}
  \begin{equation}
  \partial_i B^i\ =\ 0, \qquad\quad  \frac{\partial B^i}{\partial t}+e^{ijk} \partial_j E_k\ =\ 0,
  \end{equation}
  \begin{equation}
  \partial_i D^i\ =\ \rho, \qquad  -\frac{\partial D^i}{\partial t}+e^{ijk} \partial_j H_k\ =\ 0,
  \end{equation}
\end{subequations}
and their 4-D covariant forms are
\begin{subequations}\label{Maxwelleqs4Dflat}
  \begin{equation}
  \partial_\lambda F_{\mu\nu}+\partial_\nu F_{\lambda\mu}+\partial_\mu F_{\nu\lambda}\ =\ 0,
  \end{equation}
  \begin{equation}
  \partial_\alpha G^{\alpha\beta}\ =\ J^\beta.
  \end{equation}
\end{subequations}
The four 3-D electric and magnetic vectors, $E_i$, $H_i$, $D^i$ and $B^i$ are defined as
\begin{equation}\label{relation}
  E_i \ = \ F_{0i},\quad B^i \ = \ -\frac{1}{2c}e^{ijk}F_{jk},\quad D^i \ = \ -\frac{1}{c}G^{0i},\quad H_i \ = \ -\frac{1}{2}e_{ijk}G^{jk}.\\[1.5pt]
\end{equation}
Here we use Greek indices $\alpha, \beta, \cdots$ to run from 0 to 3 for time and spatial parts, and stipulate the sign of metric takes the form $(\,-,\ +,\ +,\ +\,)$. For vacuum, $ G^{\alpha\beta}\ =\ c\varepsilon_0F^{\alpha\beta}. $

 In curved space time, 4-D microscopic Maxwell equations which have the same form as the macroscopic Maxwell's equations in vacuum become\cite{Misner1973Freeman, Landau1995Oxford}
\begin{subequations}
  \begin{equation}
  \partial_\lambda F_{\mu\nu}+\partial_\nu F_{\lambda\mu}+\partial_\mu F_{\nu\lambda}\ =\ 0,
  \end{equation}
  \begin{equation}
  c\varepsilon_0\nabla_\alpha F^{\alpha\beta}\ =\ \frac{c\varepsilon_0}{\sqrt{-g}}\partial_{\alpha}(\sqrt{-g}F^{\alpha\beta}) \ =\ \ J^\beta,
  \end{equation}
\end{subequations}
where $\nabla_\alpha$ is 4-D covariant derivative, $g$ is the determinant of the spacetime metric $g_{\mu\nu}$.

If the excitation tensor $G^{\alpha\beta}$ is defined in the form \cite{Schurig2006OE, Leonhardt2009PO, Post1962Wiley, Plebanski1960PR}
\begin{equation}\label{FtoGMin}
  G^{\alpha\beta}\ =\ c\varepsilon_0\sqrt{-g}F^{\alpha\beta},
\end{equation}
and the definitions of 3-D electric and magnetic vectors are retained for flat Minkowski spacetime as in Eqs. (\ref{relation}), both 3-D and 4-D Maxwell's equations keep the form of flat spacetime in Eqs. (\ref{Maxwelleqs3Dflat}) and Eqs. (\ref{Maxwelleqs4Dflat}), except the 4-current $J^\beta$, charge density $\rho$, and 3-current $j^i$ should be replaced with the reduced forms $\hat{J}^{\beta}=\sqrt{-g} J^\beta,\ \hat{\rho}=\sqrt{\gamma}\rho,\ \hat{j}^i=\sqrt{\gamma}j^i$ respectively \cite{Leonhardt2009PO}, where $\gamma$ denotes the determinate of spacial metric $\gamma_{ij}\ =\ g_{ij}-g_{0i}g_{0j}/g_{00}$. We name this set of excitation tensor and 3-D electric and magnetic vectors as Minkowski's definition.

The constitutive relations in Minkowski's definition are \cite{Leonhardt2006NJP, Leonhardt2009PO, Plebanski1960PR}
\begin{subequations}
  \begin{equation}
  D^i\ =\ \varepsilon_0\sqrt{\frac{\gamma}{-g_{00}}}\gamma^{ij}E_j+\frac{1}{cg_{00}}e^{ijk}g_{0j}H_{k},
  \end{equation}
  \begin{equation}
  B^i\ =\ \mu_0\sqrt{\frac{\gamma}{-g_{00}}}\gamma^{ij}H^j-\frac{1}{cg_{00}}e^{ijk}g_{0j}E_k.,
  \end{equation}
\end{subequations}
Thus, the permittivity and permeability satisfy
$\varepsilon^{ij}=\varepsilon_0\sqrt{-\gamma/g_{00}}\gamma^{ij},\quad\mu^{ij}=\mu_0\sqrt{-\gamma/g_{00}}\gamma^{ij}$,
and magnetoelectric coupling tensor is $\kappa^{ij}=e^{ijk}g_{0j}/(cg_{00})$. If $g_{\mu\nu}$ satisfies $g_{00}=-1, g_{0i}=0$, then magnetoelectric coupling vanishes, while $\varepsilon^{ij}$ and $\mu^{ij}$ reduce to Eqs. (\ref{constitutive parameters in Minkowski}).

Another definition of 3-D electric and magnetic vectors appears in Landau's book \cite{Landau1995Oxford}, therefore, we call this set as Landau's definitions. In Landau's definition, we put a bar on each quantity, and the excitation tensor retains its formula in flat spacetime $\bar{G}^{\alpha\beta}\ =\ c\varepsilon_0F^{\alpha\beta}$, and the four electric and magnetic vectors are redefined as \cite{Landau1995Oxford, Bergamin2008PRA}
\begin{equation}\label{relationLandau}
  \bar{E}_i \ = \ F_{0i},\quad \bar{B}^i \ = \ -\frac{1}{2c\sqrt{\gamma}}e^{ijk}F_{jk},\quad \bar{D}^i \ = \ -\frac{\sqrt{-g_{00}}}{c}\bar{G}^{0i},\quad \bar{H}_i \ = \ -\frac{\sqrt{-g}}{2}e_{ijk}\bar{G}^{jk}.\\[1.5pt]
\end{equation}
Under this set of definitions, 4-D Maxwell's equations have the form of
\begin{subequations}\label{Maxwelleqs4DLandau}
  \begin{equation}
  \partial_\lambda F_{\mu\nu}+\partial_\nu F_{\lambda\mu}+\partial_\mu F_{\nu\lambda}\ =\ 0,
  \end{equation}
  \begin{equation}
  \nabla_\alpha \bar{G}^{\alpha\beta}\ =\ J^\beta.
  \end{equation}
\end{subequations}
The corresponding 3-D Maxwell's equations are
\begin{subequations}\label{Maxwelleqs3DLandau}
  \begin{equation}
  \frac{1}{\sqrt{\gamma}}\partial_i(\sqrt{\gamma}\bar{B}^i)\ =\ 0, \qquad\quad \frac{1}{\sqrt{\gamma}}\frac{\partial }{\partial t}(\sqrt{\gamma}\bar{B}^i)+\frac{1}{\sqrt{\gamma}}e^{ijk}\partial_j\bar{E}_k\ =\ 0,\\
  \end{equation}
  \begin{equation}
  \frac{1}{\sqrt{\gamma}}\frac{\partial}{\partial x^i}(\sqrt{\gamma}\bar{D}^i)\ =\ \rho, \qquad -\frac{1}{\sqrt{\gamma}}\frac{\partial }{\partial t}(\sqrt{\gamma}\bar{D}^i)+\frac{1}{\sqrt{\gamma}}e^{ijk}\partial_j\bar{H}_k\ =\ J^i.
  \end{equation}
\end{subequations}

Comparing Eqs. (\ref{relation}) with Eqs. (\ref{relationLandau}), 
we have $G^{\alpha\beta}=\sqrt{-g}\bar{G}^{\alpha\beta}$, and the transformations of other quantities between the two definitions  are identical with Eqs. (\ref{relationin3D}).

Assume that an observer whose 4-velocity is $u^\alpha$ in a curved spacetime endowed with an electromagnetic field $F_{\mu\nu}$. Then, questions arouse naturally. What is the measurement of electric and magnetic fields by the observer? Is the measurement equal to one of the two definitions mentioned above? Actually, there has been a mature way to answer this question. In the curved spacetime, electric and magnetic field vectors are defined respected with the frame $u^\alpha$ as \cite{Felice1990Cambridge}
\begin{equation}\label{EHobserver}
  \tilde{E}_{\alpha}\ =\ -F_{\alpha\beta}u^{\beta},\qquad \tilde{H}_{\alpha}\ =\ ^{\ast}\bar{G}_{\alpha\beta}u^{\beta}=\frac{1}{2}\epsilon_{\alpha\beta\mu\nu}\bar{G}^{\mu\nu}u^{\beta},
\end{equation}
where $^{\ast}\bar{G}_{\alpha\beta}=\epsilon_{\alpha\beta\mu\nu}\bar{G}^{\mu\nu}/2$ is the dual of $\bar{G}^{\mu\nu}$. Here, we put a tilde on electric and magnetic vectors to discriminate its representation under Minkowski's definition and Landau's definition, however, the definition of excitation tensor retains its formula from Landau's definition  $\bar{G}^{\alpha\beta}\ =\ c\varepsilon_0F^{\alpha\beta}$, which guarantees that $\bar{G}$ is a tensor of 4-dimension.

In terms of the definitions of Eqs. (\ref{EHobserver}), $\tilde{E}_{\alpha}u^{\alpha}=\tilde{H}_{\alpha}u^{\alpha}=0$, which evinces that $\tilde{E}$ and $\tilde{H}$ are both space-like vectors for the observer $u^\alpha$. In addition, if we simply let electric displacement $\tilde{D}^\alpha=\varepsilon_0\tilde{E}^\alpha$ and magnetic induction $\tilde{B}^\alpha=\mu_0\tilde{H}^\alpha$,  {\it i.e.} what are made in vacuum of flat spacetime, the components of  $\tilde{E}$, $\tilde{H}$, $\tilde{D}$, and $\tilde{B}$ in the tetrad $e_{\scriptscriptstyle{\langle}\scriptstyle{\mu}\scriptscriptstyle{\rangle}}^{\ \alpha}$ of the observer, where  $e_{\scriptscriptstyle{\langle}\scriptstyle{0}\scriptscriptstyle{\rangle}}^{\ \alpha}=u^{\alpha}$, are
\begin{equation}\label{EHobservercomponent}
  \tilde{E}_{\langle i\rangle}\ =\ F_{\langle 0i\rangle},\ \
  \tilde{H}_{\langle i\rangle}\ =\ -\frac{1}{2}e_{ijk}\bar{G}^{\langle jk\rangle},\ \
  \tilde{D}^{\langle i\rangle}\ =\ -\frac{1}{c}\bar{G}_{\langle 0i\rangle},\ \
  \tilde{B}^{\langle i\rangle}\ =\ -\frac{c}{2}e^{ijk}F_{\langle jk\rangle}.
\end{equation}
As we see, the components in tetrad satisfy the original relation of flat spacetime (\ref{relation}). Furthermore, the electromagnetic tensors admit the following decomposition \cite{Felice1990Cambridge, Sonego1998JMP, Tsagas2005CQG}:
\begin{subequations}\label{relationobserver}
  \begin{equation}
  F_{\mu\nu}\ =\ (\tilde{E}_{\mu}u_\nu-\tilde{E}_{\nu}u_\mu)-c\epsilon_{\mu\nu\alpha\beta}u^\alpha \tilde{B}^\beta,
  \end{equation}
  \begin{equation}
  \bar{G}^{\mu\nu}\ =\ c(\tilde{D}^{\mu}u^\nu-\tilde{D}^{\nu}u^\mu)-\epsilon^{\mu\nu\alpha\beta}u_\alpha \tilde{H}_\beta,
  \end{equation}
\end{subequations}
and a further study reveals the form of Maxwell's equations expressed with $\tilde{E}$ and $\tilde{H}$ \cite{Sonego1998JMP, Tsagas2005CQG}.

Here, we are interested in the condition that the observer is comoving. For a comoving observer, $u^\alpha=(1/\sqrt{-g_{00}},\ 0,\ 0,\ 0)$, and the transverse projector $h_{\mu\nu}=g_{\mu\nu}+u_\mu u_\nu$ reduces to the spatial metric $\gamma_{ij}$ mentioned earlier. In this case, the relations (\ref{relationobserver}) lead to the result
\begin{equation}
\left\{
\begin{array}{rcl}\label{relationcomovingobserver}
  F_{0i} & = & \sqrt{-g_{00}}\tilde{E}_i,\\[4pt]
  F_{jk} & = & -c\sqrt{\gamma}e_{ijk}\tilde{B}^i +\frac{2}{\sqrt{-g_{00}}}\tilde{E}_{[j}g_{k]0},\\[5pt]
  \bar{G}^{0i} & = & -\frac{c}{\sqrt{-g_{00}}}\tilde{D}^i+\frac{1}{\sqrt{g_{00}g}}e^{ijk}g_{j0}\tilde{H}_k,\\[6pt]
  \bar{G}^{jk} & = & -\frac{1}{\sqrt{\gamma}}e^{ijk}\tilde{H}_i.\\[5pt]
  \end{array}
\right.
\end{equation}
It seems that the measurement of electric and magnetic fields by an observer  conforms neither to Minkowski's definition nor to Landau's definition. However, if we set $g_{00}=-1, g_{0i}=0$, the reduced relations Eqs. (\ref{relationcomovingobserver}) are identical with the reduced relations Eqs. (\ref{relationLandau}) under Landau's definition. So, if we use artificial materials to simulate the behavior of electromagnetic fields in curved spacetime based on the form equivalence of Maxwell's equations, it should be heeded that the distinction between the electric and magnetic fields measured in materials and measured by the observer moving in real curved spacetime.

\section*{Appendix B:\quad Electromagnetic fields in Cylindrical cloak}
\stepcounter{appendix}
\setcounter{equation}{0}
In this portion, we would only consider TE polarized wave. Substituting the constitutive parameters Eq. (\ref{constitutiveparametersincylindricalcloak})into wave equation (\ref{wave equation}), we get the general scalar wave equation for all transformation $f(r)$ which is exactly the Helmholtz equation in original {\bf S$'$} coordinate system of virtual space:
\begin{equation}\label{general scalar wave equation}
  \frac{1}{f}\,\frac{\partial}{\partial f}\left( f\frac{\partial {E_{\scriptscriptstyle(\mathrm{P})}}_z}{\partial f} \right)\ +\ \frac{1}{f^2}\,\frac{\partial^2 {E_{\scriptscriptstyle(\mathrm{P})}}_z}{\partial \theta^2}\ +\ k^2\, {E_{\scriptscriptstyle(\mathrm{P})}}_z\ =\ 0.
\end{equation}
Through separating variables, electromagnetic fields can be represented by series
\begin{flalign}\label{general scalar wave equation}
  \vec{E}_{\scriptscriptstyle(\mathrm{P})}\ =\ &\ \sum_{-\infty}^{\infty}\ \left[\ a_n\,J_n(x)\ +\ b_n\,N_n(x)\ \right]\;\me ^{\mi n\theta}\ \hat{e}_z,
\end{flalign}
where $J_n(x)$ and $N_n(x)$ are n-order Bessel function and n-order Neumann function respectively. These expressions are available in every area, however $x$ takes different forms in each area. Outside the cloak ($r>b$), $x=k_0r$ is in the expressions of incident wave $\vec{E}_{\scriptscriptstyle(\mathrm{P})}^{\mathrm{in}}$ and scattering wave $\vec{E}_{\scriptscriptstyle(\mathrm{P})}^{\mathrm{sc}}$. In the cloak layer ($a<r<b$),  $\vec{E}_{\scriptscriptstyle(\mathrm{P})}^\mathrm{c}$ takes $x=k_0f(r)$. Inside the internal hidden area ($r<a$), $x=k_1r$ is in the expression of $\vec{E}_{\scriptscriptstyle(\mathrm{P})}^{\mathrm{int}}$, where $k_1=\omega\sqrt{\varepsilon_1\mu_1}$ and $\varepsilon_1,\ \mu_1$ are the material properties of the hidden medium. The incident wave $\vec{E}_{\scriptscriptstyle(\mathrm{P})}^{\mathrm{in}}=E_0\me^{\mi k_0r\cos\theta}\hat{e}_z$ can be expanded as
\begin{equation}\label{incident wave}
  \vec{E}_{\scriptscriptstyle(\mathrm{P})}^{\mathrm{in}}\ =\ E_0\sum_{-\infty}^{\infty}\ \mi^n\;J_n(k_0r)\;\me^{\mi n\theta}.
\end{equation}
Because $\vec{E}^\mathrm{sc}$ should tend to an outgoing wave as $r$ approaches to infinity, it is more convenient to expand scattering wave with Hankel function of first kind:
\begin{equation}\label{scattering wave}
  \vec{E}_{\scriptscriptstyle(\mathrm{P})}^{\mathrm{sc}}\ =\ \sum_{-\infty}^{\infty}\ a_n^{\mathrm{sc}}H_n^{\scriptscriptstyle(1)}(k_0r)\,\me^{\mi n\theta}.
\end{equation}
In addition, $b_n^{\mathrm{int}}=0$ means that fields should be finite when $r\rightarrow0$. Then applying the continuity of $E_z$ and $H_\theta$ at inner and outer surface of cloak, we have following equations:
\begin{subequations}\label{joining equations}
\begin{flalign}
  \label{joining equations1} a_n^{\mathrm{int}}\,J_n(k_1a) \ &=\ a_n^{\mathrm{c}}\,J_n\big(k_0f(a)\big)\ +\ b_n^{\mathrm{c}}\,N_n\big(k_0f(a)\big),\\
  \label{joining equations2} a_n^{\mathrm{int}}\,\frac{\mu_0}{\mu_1}k_1J_n(k_1a) \ &=\ \frac{k_0f(a)}{a}\left[a_n^{\mathrm{c}}\,J_n'\big(k_0f(a)\big)\ +\ b_n^{\mathrm{c}}\,N_n'\big(k_0f(a)\big)\right],\\
  E_0\mi^n\,J_n(k_0b)\ +\ a_n^{\mathrm{sc}}\,H_n^{\scriptscriptstyle(1)}(k_0b)\ &=\ a_n^{\mathrm{c}}\,J_n\big(k_0f(b)\big)\ +\ b_n^{\mathrm{c}}\,N_n\big(k_0f(b)\big),\\
  E_0\mi^n\,J_n'(k_0b)\ +\ a_n^{\mathrm{sc}}\,H_n^{\scriptscriptstyle(1)\prime}(k_0b)\ &=\ \frac{f(b)}{b}\left[a_n^{\mathrm{c}}\,J_n'\big(k_0f(b)\big)\ -\ b_n^{\mathrm{c}}\,N_n'\big(k_0f(b)\big)\right],
\end{flalign}
\end{subequations}
Solving the group of equations, we obtain
\begin{subequations}\label{coeficients}
\begin{flalign}
  \begin{split}
  a_n^{\mathrm{c}}\ =\ &E_0\,\mi^n\,\left[J_n(k_0b)H_n^{\scriptscriptstyle(1)\prime}(k_0b)\ -\ J_n'(k_0b)H_n^{\scriptscriptstyle(1)}(k_0b)\right]/\Big\{H_n^{\scriptscriptstyle(1)\prime}(k_0b)\left[ J_n\big(k_0f(b)\big) \right.\\
  &\left.\left.+ A_nN_n\big(k_0f(b)\big) \right]-\frac{f(b)}{b}H_n^{\scriptscriptstyle(1)}(k_0b)\left[ J_n'\big(k_0f(b)\big)+A_nN_n'\big(k_0f(b)\big) \right]\right\},
  \end{split}\\
  b_n^{\mathrm{c}}\ =\ &\frac{\mu_0ak_1J_n\big(k_0f(a)\big)J_n'(k_1a)\ -\ \mu_1k_0f(a)J_n'\big(k_0f(a)\big)J_n(k_1a)}{\mu_1k_0f(a)N_n'\big(k_0f(a)\big)J_n(k_1a)-\mu_0ak_1N_n\big(k_0f(a)\big)J_n'(k_1a)}\ a_n^{\mathrm{c}}\ =\ A_n\;a_n^{\mathrm{c}},\\
  a_n^{\mathrm{int}}\ =\ &\frac{1}{J_n(k_1a)}\;\left[\,J_n\big(k_0f(a)\big)\ +\ A_n\,N_n\big(k_0f(a)\big)\, \right]\ a_n^{\mathrm{c}},\\
  a_n^{\mathrm{sc}}\ =\ &\frac{1}{H_n^{\scriptscriptstyle(1)\prime}(k_0b)}\left\{ \frac{f(b)}{b}\left[\,J_n'\big(k_0f(b)\big)\ -\ A_n\,N_n'\big(k_0f(b)\big)\right]a_n^{\mathrm{c}}\ -E_0\mi^n\,J_n'(k_0b) \right\}.
\end{flalign}
\end{subequations}
Substituting invisibility condition $f(a)=0,\ f(b)=b$, we have $  a_n^{\mathrm{c}} =E_0\;\mi^n$, $b_n^{\mathrm{c}}  = 0\ (\ i.e.\ A_n=0\ )$, therefore $a_n^{\mathrm{sc}}\ =\ 0$ which indicates no scattering. However, when calculating $a_n^{\mathrm{int}}$, there is a term $A_n\cdot N_n\big(k_0f(a)\big)$, in which $A_n\rightarrow0$ while $N_n\big(k_0f(a)\big)\rightarrow\infty$ as $f(a)\rightarrow0$. A meticulous calculation gives the result $a_n^{\mathrm{int}}\rightarrow0$, so waves can not spread into the hidden area. Finally, we get the expressions of electromagnetic fields in the whole space as shown in Eqs.(\ref{FieldsinCylindricalcloak}).

A detailed derivation should be identified. If we substitute the
invisibility condition $f(a)=0,\ f(b)=b$ at the beginning, the
fields in the cloak layer must have no terms of Neumann function,
which means $b_n^{\mathrm{c}}  = 0$ , since
$N_n\big(k_0f(a)\big)\rightarrow\infty$ is against the
requirement that fields should be finite. Therefore, the last terms
vanish in Eqs. (\ref{joining equations1}) and (\ref{joining
equations2}). In this case, the Eq. (\ref{joining equations2}) still
leads to  $a_n^{\mathrm{int}}=0$, yet the result can not satisfy Eq.
(\ref{joining equations1}) when $n=0$. In the more general
derivation above, when $n=0$, though $b_n^{\mathrm{c}} \rightarrow
0$, the product $b_n^{\mathrm{c}}\cdot N_n\big(k_0f(a)\big)$ is
towards to a finite quantity $-E_0\mi^n/J_n(k_1a)$, and it
counterbalances the equality (\ref{joining equations1}), however,
the product only exists at the inner surface, so it no longer has
the meaning of fields but represents the magnetic surface current
flowing along the $z$ axis \cite{Zhang2007PRB}.

\section*{Appendix C:\quad Eikonal equation of different forms}
\stepcounter{appendix}
\setcounter{equation}{0}

Wave equation (\ref{wave equation}) can be written as the components form
\begin{equation}
   \left[  \epsilon_{ipj}k^p{(\bar{n}^{\scriptscriptstyle-1})}^{jk}\epsilon_{kql}k^q+\left(\frac{\omega}{c}\right)^2\bar{n}_{il}  \right]E^l\ =\ 0.
\end{equation}
The condition which protects $E^l$ from trivial solution is
\begin{equation}
   \det\left[  \epsilon_{ipj}k^p{(\bar{n}^{\scriptscriptstyle-1})}^{jk}\epsilon_{kql}k^q+\left(\frac{\omega}{c}\right)^2\bar{n}_{il}  \right]\ =\ 0.
\end{equation}
Expanding the determinant
\begin{equation}\label{condition of untrival solution}
  \begin{split}
   & \det\left[  \epsilon_{ipj}k^p{(\bar{n}^{\scriptscriptstyle-1})}^{jk}\epsilon_{kql}k^q+\left(\frac{\omega}{c}\right)^2\bar{n}_{il}  \right]\\
   =\ &e^{i_1i_2i_3}\cdot\prod^3_{\alpha=1}\left[  \epsilon_{i_\alpha p_\alpha j_\alpha }k^{p_\alpha}{(\bar{n}^{\scriptscriptstyle-1})}^{j_\alpha k_\alpha}\epsilon_{k_\alpha q_\alpha \alpha}k^{q_\alpha}+\left(\frac{\omega}{c}\right)^2\bar{n}_{i_\alpha \alpha}  \right]\\
   =\ &\sqrt{\gamma}\epsilon^{i_1i_2i_3}\left[ \prod^3_{\alpha=1}M_{i_\alpha\alpha}+\left(\frac{\omega}{c}\right)^2\sum^3_{\alpha=1}\bar{n}_{i_\alpha\alpha}M_{i_\alpha(\alpha+1)}M_{i_\alpha(\alpha+2)} \right. \\
    & \left. + \left(\frac{\omega}{c}\right)^4\sum^3_{\alpha=1}M_{i_\alpha\alpha}\bar{n}_{i_{\alpha+1}(\alpha+1)}\bar{n}_{i_{\alpha+2
   }(\alpha+2)} + \left(\frac{\omega}{c}\right)^6\prod^3_{\alpha=1}\bar{n}_{i_\alpha \alpha}
    \right]
  \end{split}
\end{equation}
where $M_{il}=\epsilon_{ipj}N^{pj}_l=\epsilon_{ipj}k^p{(\bar{n}^{\scriptscriptstyle-1})}^{jk}\epsilon_{kql}k^q$, and the indices of $\epsilon_{i_\alpha j_\alpha k_\alpha}$ satisfy $\alpha \mod\ 3$. Calculating each term in the expansion, we have
\begin{flalign}
 \mbox{1st term:} & &
    & \sqrt{\gamma}\epsilon^{i_1i_2i_3} \prod^3_{\alpha=1}M_{i_\alpha\alpha}\ = \  \sqrt{\gamma}\epsilon^{i_1i_2i_3}\epsilon_{i_1p_1j_1}\epsilon_{i_2p_2j_2}\epsilon_{i_3p_3j_3}N_1^{p_1j_1}N_2^{p_2j_2}N_3^{p_3j_3}&&\notag\\
  && =\ & \sqrt{\gamma}\left( \epsilon_{p_1p_2j_2}\epsilon_{j_1p_3j_3}-\epsilon_{j_1p_2j_2}\epsilon_{p_1p_3j_3} \right)N_1^{p_1j_1}N_2^{p_2j_2}N_3^{p_3j_3}=\  0,& &
\end{flalign}
\begin{flalign}
  \mbox{2nd term:} & &
     &\sqrt{\gamma}\left(\frac{\omega}{c}\right)^2\sum^3_{\alpha=1}\epsilon^{i_1i_2i_3}\bar{n}_{i_\alpha\alpha}M_{i_\alpha(\alpha+1)}M_{i_\alpha(\alpha+2)}&&\notag\\
  && =\ & \sqrt{\gamma}\left(\frac{\omega}{c}\right)^2\sum^3_{\alpha=1}\epsilon^{i_{\alpha}i_{\alpha+1}i_{\alpha+2}}\bar{n}_{i_\alpha\alpha}M_{i_\alpha(\alpha+1)}M_{i_\alpha(\alpha+2)}&&\notag\\
  && =\ & \sqrt{\gamma}\left(\frac{\omega}{c}\right)^2\sum^3_{\alpha=1}\bar{n}_{i_\alpha \alpha}\epsilon_{p_{\alpha+2}j_{\alpha+1}j_{\alpha+2}}N^{p_{\alpha+1}j_{\alpha+1}}_{\alpha+1}N^{p_{\alpha+2}j_{\alpha+2}}_{\alpha+2},&& \notag
\end{flalign}
\begin{flalign}
  \hspace{35pt}&&&\hspace{-40pt}\mbox{because}&&\notag\\
    &&& \mbox{2nd term}\cdot\det\left(\bar{n}_{ij}\right)&&\notag\\
   && =\ & \sqrt{\gamma}\left(\frac{\omega}{c}\right)^2\sum^3_{\alpha=1}\bar{n}_{i_\alpha \alpha}\epsilon_{p_{\alpha+2}j_{\alpha+1}j_{\alpha+2}}\cdot\left[ k^{i_\alpha}{(\bar{n}^{\scriptscriptstyle-1})}^{j_{\alpha+1}k_{\alpha+1}}\epsilon_{k_{\alpha+1}q_{\alpha+1}(\alpha+1)}k^{q_{\alpha+1}} \right]&&\notag\\
   && & \cdot\left[ k^{p_{\alpha+2}}{(\bar{n}^{\scriptscriptstyle-1})}^{j_{\alpha+2}k_{\alpha+2}}\epsilon_{k_{\alpha+2}q_{\alpha+2}(\alpha+2)}k^{q_{\alpha+2}} \right]\cdot \left[ \sqrt{\gamma}\epsilon^{t_1t_2t_3}\bar{n}_{t_11}\bar{n}_{t_22}\bar{n}_{t_33} \right]&&\notag\\
   && =\ & \gamma\left(\frac{\omega}{c}\right)^2\left[  \bar{n}_{i_\alpha \alpha}\bar{n}_{t_\beta \beta}k^{i_\alpha}k^{t_\beta}\gamma k^\alpha k^\beta \right]\ =\  \left(\gamma\frac{\omega}{c} \bar{n}_{ij}k^ik^j \right)^2,&&\notag\\
   &&&\hspace{-40pt}\mbox{therefore we have}\mspace{10mu}&&\notag\\
   &&&\hspace{35pt}\mbox{2nd term} \ = \ \left(\gamma\frac{\omega}{c} \bar{n}_{ij}k^ik^j \right)^2/\det\left(\bar{n}_{ij}\right),&&
\end{flalign}
\begin{flalign}
 \mbox{3rd term:} & &
    & \sqrt{\gamma}\left(\frac{\omega}{c}\right)^4\sum^3_{\alpha=1}\epsilon^{i_1i_2i_3}M_{i_\alpha\alpha}\bar{n}_{i_{\alpha+1}(\alpha+1)}\bar{n}_{i_{\alpha+2
   }(\alpha+2)}&&\notag&&\notag\\
  && =\ &  \sqrt{\gamma}\left(\frac{\omega}{c}\right)^4\sum^3_{\alpha=1}\epsilon^{i_{\alpha}i_{\alpha+1}i_{\alpha+2}}\epsilon_{i_\alpha p_\alpha j_\alpha}\bar{n}_{i_{\alpha+1}(\alpha+1)}\bar{n}_{i_{\alpha+2
   }(\alpha+2)}N^{p_\alpha j_\alpha}_\alpha&&\notag\\
  && =\ & \sqrt{\gamma}\left(\frac{\omega}{c}\right)^4\sum^3_{\alpha=1}\left[ \bar{n}_{p_\alpha(\alpha+1)}\epsilon_{q_\alpha \alpha (\alpha+2)}-\bar{n}_{p_\alpha(\alpha+2)}\epsilon_{q_\alpha \alpha (\alpha+1)}\right]k^{p_\alpha}k^{q_\alpha}&&\notag\\
  && =\ & -2\gamma\left(\frac{\omega}{c}\right)^4\bar{n}_{ij}k^ik^j,& &
\end{flalign}
\begin{flalign}
 \mbox{4th term:} & &
    \sqrt{\gamma}\left(\frac{\omega}{c}\right)^6\epsilon^{i_1i_2i_3}\prod^3_{\alpha=1}\bar{n}_{i_\alpha \alpha}\ =\ \left(\frac{\omega}{c}\right)^6\det\left( \bar{n}_{ij} \right).& &
\end{flalign}
Substituting them into Eq. (\ref{condition of untrival solution}), we have
\begin{equation}
   \det\left[  \epsilon_{ipj}k^p{(\bar{n}^{\scriptscriptstyle-1})}^{jk}\epsilon_{kql}k^q+\left(\frac{\omega}{c}\right)^2\bar{n}_{il}  \right]\ =\ \frac{(\gamma\omega/c)^2}{\det\left( \bar{n}_{ij} \right)}\left[ \bar{n}_{ij}k^ik^j-\frac{1}{\gamma}\left(\frac{\omega}{c}\right)^2\det\left( \bar{n}_{ij} \right) \right]^2\ =\ 0.
\end{equation}
This is exactly the wave-vector eikonal equation (\ref{dispersion relation})  expressed by Landau's definition. The ray-vector eikonal equation (\ref{dual dispersion relation}) can be simply obtained through duality rules. In terms of Eq. (\ref{relationin3D2}) and Eq. (\ref{components in unit bases}), we can get the two kinds of eikonal equation written by Minkowski's components and anholonomic components.

Under Minkowski's definition, wave-vector eikonal equation is
\begin{equation}
   n^{ij}k_ik_j-\left(\frac{\omega}{c}\right)^2\det\left( n^{ij} \right)\ =\ n_{ij}k^ik^j-\frac{1}{\gamma^2}\left(\frac{\omega}{c}\right)^2\det\left( n_{ij} \right)\ =\ 0,
\end{equation}
and ray-vector eikonal equation is
\begin{equation}
   {(n^{\scriptscriptstyle-1})}^{ij}s_is_j-\gamma^2\left(\frac{c}{\omega}\right)^2\det\left[ (n^{\scriptscriptstyle-1})^{ij} \right]\ =\ {(n^{\scriptscriptstyle-1})}_{ij}s^is^j-\left(\frac{c}{\omega}\right)^2\det\left[ (n^{\scriptscriptstyle-1})_{ij} \right]\ =\ 0.
\end{equation}
The wave-vector eikonal equation expressed by the components in anholonomic unit basis is
\begin{equation}
   {(n^{\scriptscriptstyle-1})}_{\langle ij \rangle}k_{\langle i \rangle}k_{\langle j \rangle}-\left(\frac{\omega}{c}\right)^2\det\left[ (n^{\scriptscriptstyle-1})_{\langle ij \rangle} \right]\ =\ 0,
\end{equation}
and the ray-vector eikonal equation is
\begin{equation}
   {(n^{\scriptscriptstyle-1})}_{\langle ij \rangle}s_{\langle i \rangle}s_{\langle j \rangle}-\left(\frac{c}{\omega}\right)^2\det\left[ (n^{\scriptscriptstyle-1})_{\langle ij \rangle} \right]\ =\ 0.
\end{equation}


\bibliography{Analytical_form_of_light-ray_tracing_in_invisibility_cloaks}

\end{document}